\documentstyle[aps,floats,psfig]{revtex}

\begin{document}
\tighten
\draft

\twocolumn[\hsize\textwidth\columnwidth\hsize\csname
@twocolumnfalse\endcsname
\begin{flushright}
CERN-TH/2000-140\\
SPhT 00/45\\
\end{flushright}

\title{Exhaustive Study of Cosmic Microwave Background Anisotropies in 
Quintessential Scenarios}
\author{Philippe Brax}
\address{CERN, Theory division, Gen\`eve, Switzerland, \\
(on leave from SPhT-Saclay CEA 
F-91191 Gif-sur-Yvette, France.) \\
e-mail: Philippe.Brax@cern.ch }
\author{J\'er\^ome Martin and Alain Riazuelo}
\address{DARC, Observatoire de Paris, \\
UMR 8629 CNRS, F-92195 Meudon Cedex, France. \\
e-mail: martin@edelweiss.obspm.fr, Alain.Riazuelo@obspm.fr}

\maketitle

\begin{abstract}
Recent high precision measurements of the CMB anisotropies performed
by the BOOMERanG and MAXIMA-1 experiments provide an unmatched set of
data allowing to probe different cosmological models. Among these
scenarios, motivated by the recent measurements of the luminosity
distance versus redshift relation for type Ia supernovae, is the
quintessence hypothesis. It consists in assuming that the acceleration
of the Universe is due to a scalar field whose final evolution is
insensitive to the initial conditions. Within this framework we
investigate the cosmological perturbations for two well-motivated
potentials: the Ratra-Peebles and the SUGRA tracking potentials. We
show that the solutions of the perturbed equations possess an
attractor and that, as a consequence, the insensitivity to the initial
conditions is preserved at the perturbed level. Then, we study the
predictions of these two models for structure formation and CMB
anisotropies and investigate the general features of the multipole
moments in the presence of quintessence. We also compare the CMB
multipoles calculated with the help of a full Boltzmann code with the
BOOMERanG and MAXIMA-1 data. We pay special attention to the location
of the second peak and demonstrate that it significantly differs from
the location obtained in the cosmological constant case. Finally, we
argue that the SUGRA potential is compatible with all the recent data
with a standard values of the cosmological parameters. In 
particular, it fits the MAXIMA-1 data better than a cosmological
constant or the Ratra-Peebles potential.
\end{abstract}

\narrowtext
\vspace{1 cm}]

\newcommand{\ETAL}{{\it et al}}
\newcommand{\IE}{i.e.}

\newcommand{\BAR}{{\mathrm b}}
\newcommand{\CDM}{{\mathrm cdm}}
\newcommand{\MAT}{{\mathrm m}}
\newcommand{\RAD}{{\mathrm r}}

\newcommand{\BG}{{\mathrm B}}

\newcommand{\CRIT}{{\mathrm c}}
\newcommand{\PL}{{\mathrm Pl}}
\newcommand{\MAX}{{\mathrm max}}

\newcommand{\SUSY}{{\mathrm SUSY}}
\newcommand{\SUGRA}{{\mathrm SUGRA}}
\newcommand{\RP}{{\mathrm RP}}

\newcommand{\QUINT}{{\mathrm Q}}

\newcommand{\SCAL}{{\mathrm S}}
\newcommand{\TENS}{{\mathrm T}}
\newcommand{\ddd}{{\mathrm d}}
\newcommand{\Hconf}{{\mathcal H}}
\newcommand{\vson}{{c^2_{\mathrm s}}}

\newcommand{\UUNIT}[2]{
\,{\rm {#1}}^{#2}}

\section{Introduction}

Recent measurements of the luminosity distance versus redshift
relation for type Ia supernovae~\cite{SNIa}, if confirmed, are
compatible with an expanding (accelerating) universe driven by a new
type of matter whose equation of state $p = \omega \rho$ is
characterized by a negative $\omega$. One of 
the possible explanations is the
existence of a non-zero vacuum energy, \IE{} a ``cosmological
constant''. Another pragmatic possibility which has been proposed is
to assume the existence of a yet unknown mechanism guarenteeing that
the true cosmological constant vanishes, the remaining energy density
being then due to the presence of a scalar field, the quintessence
field, almost decoupled from ordinary
matter~\cite{RP,FJ1,ZWS,SWZ}. The main difference between a
quintessence fluid and a cosmological constant comes from their
equation of state where $\omega_\Lambda = -1$ for a cosmological
constant and $- 1 \le \omega_\QUINT \le 0$ for the quintessence fluid.
\par
One of the puzzles in the interpretation of these data is the
extremely small value of the energy density due to the new form of
matter. From the point of view of particle physics a vanishing value
for the cosmological constant is one of the major
challenges~\cite{Wein}. At present there is no known mechanism which
prevents the vacuum energy from picking large values due to radiative
corrections and one expects typically a contribution equal to $(\hbar
c / 2) \int \ddd {\bf k} k /(2 \pi)^3 \simeq \hbar c k_\MAX^4 /(16
\pi^2)$, where $k_\MAX$ is a cut-off which can naturally be taken as
the Planck wavenumber. This gives a contribution which is $120$ orders
of magnitude above the observed one. One possibility which is often
advocated is the presence of some global supersymmetry (SUSY) which
would guarantee that the energy of the vacuum is zero. Unfortunately
SUSY has to be broken to take into account the absence of experimental
evidence in favour of particle superpartners leading to a natural
contribution to the vacuum energy of order $M_\SUSY^4$ where
$M_\SUSY$ is the SUSY breaking scale estimated around $1
\UUNIT{TeV}{}$~\cite{Witten}. The measurement of a vacuum energy some
$60$ orders of magnitude below this expected value indicates that some
new physics must be at play here.
\par
In the quintessence hypothesis, the small vacuum energy density is due
to the rolling down of the quintessence field $Q$ along a
decreasing potential. A typical potential is the Ratra-Peebles
potential $V(Q) = \Lambda^{4 + \alpha} / Q^{\alpha}$~\cite{RP}. From
the particle physics point of view one would like to justify the
existence of the quintessence field. Several natural candidates have
been ruled out such as the axion-dilaton field~\cite{Bine}, the moduli
fields of toroidal compactifications in string theory~\cite{macorra}
and finally the meson fields of supersymmetric gauge
theories~\cite{BM2}. Nevertheless, it seems reasonnable to expect that
SUSY will play a role in the solution. Within this framework it is a
matter of fact that the quintessence field must be part of
supergravity (SUGRA) models~\cite{BM2,BM1}. This comes from the large
value $Q \simeq m_\PL$ of the field at small redshift which implies
that SUGRA corrections cannot be neglected. In~\cite{BM1} an effective
theory approach has been used to deduce the general form of
quintessence SUGRA potentials, they are of the type
\begin{equation}
\label{citepot}
 V(Q) = \frac{\Lambda^{4 + \alpha}}{Q^\alpha} e^{\kappa Q^2 / 2} ,
\end{equation}
where $\kappa \equiv 8 \pi G$, $G$ being the Newton constant, and
where the exponential factor comprises the SUGRA corrections. $\Lambda
$ and $\alpha $ are free parameters. The fine-tuning is not too severe
as for typical values $\alpha = 6$ the scale $\Lambda \simeq 10^6
\UUNIT{GeV}{}$ is compatible with high energy scales. Notice that the
SUGRA corrections become relevant towards the end of the evolution and
decouple at small $Q \ll m_\PL$. Different types of potentials can be
distinguished because they lead to different values of the equation of
state parameter. For example, for $\alpha = 11$, the Ratra-Peebles
potential is such that $\omega_\QUINT \simeq -0.29$ whereas the SUGRA
potential gives $\omega_\QUINT \simeq -0.82$~\cite{BM1} (for
$\Omega_\QUINT = 0.7$).
\par
It is also worth noticing that there exists quintessence models where
the field is non-minimally coupled with the metric. Such models induce
time-variation of the Newton constant and are therefore already
constrained, for example by observations in the solar system or by 
pulsar timing measurements~\cite{carroll,chiba}. They lead to
the same tracking behaviour, as stressed in Refs.~\cite{jpu,amendola},
as soon as the coupling term is proportional to a power of the
potential. However, some important differences occur when the field
starts dominating; for example its effective equation of state can
reach extreme values such that $\omega \simeq -3$~\cite{ur2000}. Also,
these models can lead (especially in the context of quintessential
inflation~\cite{pv}) to clear observable features in the gravitational
waves spectrum~\cite{giovannini}.
\par
In view of the numerous phenomenological successes of quintessence, it
is relevant to deduce its consequences for Cosmic Microwave Background
(CMB) anisotropies and structure formation. The aim is
two--fold. First, we have to study whether quintessence leads to
acceptable scenarios and, second, we have to learn how we could use
high-precision measurements recently obtained by the
BOOMERanG~\cite{Boo,B98a,B98b} and MAXIMA-1~\cite{MAX1a,MAX1b}
experiments or to be performed in the near future by NASA's 
Microwave Anisotropy Probe (MAP) satellite~\cite{map}, ESA's Planck
satellite~\cite{Pl} or the Sloan Digital Sky Survey (SDSS)~\cite{sloan} 
to put constraints on the quantities
characterizing quintessence like $\Omega_\QUINT$ or $\omega_\QUINT$.
The second possibility has of course already been investigated for the
cosmological constant case. For example, the fraction $\Omega_\Lambda$
of the critical density is not determined entirely from the supernovae
data. Indeed, the data from the supernovae observations are degenerate
in the plane $(\Omega_\MAT, \Omega_\Lambda)$, where $\Omega_\MAT$
is the matter (\IE{} cold dark matter plus baryons) component
preventing a clear cut determination of the fraction
$\Omega_\Lambda$. The situation changes drastically if one includes
the measurements of the CMB anisotropies~\cite{Teg} (even without the
BOOMERanG or MAXIMA-1 data). In that case, the degeneracy is removed
leading to a probable $70\%$ of the total energy density of the
universe carried by the negative pressure fluid while the remaining
$30\%$ are the matter components ensuring that $\Omega_0 = 1$ in
agreement with a spatially flat universe. This conclusion can be drawn
from the measurements of the location of the first Doppler peak. This
result has been confirmed by other
measurements~\cite{ZD,WCOS,WF}. Another use of combined data has been
to put constraints on the equation of state parameter. However, this
has been done only for constant or for very simple time-dependent
$\omega_\QUINT$~\cite{PTW,CH,Efst}.
\par
CMB anisotropies and the power spectrum are calculated with the help
of the theory of cosmological perturbations. Cosmological
perturbations in the presence of quintessence have been studied by
Ratra and Peebles but only in the tracking regime~\cite{RP}. CMB
multipoles moments and/or the power spectrum have already been
calculated for the Ratra-Peebles potential in Ref.~\cite{CDS} 
and for other models of quintessence in
Refs.~\cite{FJ2,BP,BPM,MCBW,DKS}. One important issue is to understand
whether the final evolution of the various perturbed quantities depend
on the initial conditions imposed at reheating (of the inflationary
type or not). Another way to put the same problem is the following: do
the multipole moments depend on the value of $\delta Q(\eta_{\rm i})$
and $\delta Q'(\eta_{\rm i})$ at initial time? In Ref.~\cite{CDS}, it
was noticed that the answer to this question is no but no explanations
were provided. Here, we confirm the remark of Ref.~\cite{CDS} and show
that this is due to the fact that the perturbed Einstein equations
also possess an attractor which renders the multipole moments
insensitive to the initial conditions.
\par
One of the main purpose of this article is the study of the general
properties of the multipoles moments of the CMB anisotropies in
presence of the quintessence field. We present the CMB multipole
moments for the Ratra-Peebles potential and, for the first time, for
the SUGRA tracking potential. In addition, we also display the matter
power spectrum for these two models. Recently, it has been shown by
Kamionkowski and Buchalter~\cite{KB} that the location of the second
peak in the CMB power spectrum is an efficient way of revealing some
features of the dark energy sector. Therefore, we pay special
attention to this question. In particular, in Ref.~\cite{KB}, only the
cosmological constant case was studied and it was argued that the
quintessence case (the authors refer to the Ratra-Peebles potential)
must not differ significantly from the cosmological constant case. In
the present article, we demonstrate that this is not the case and
that, as a matter of fact, quintessence leads to a different location
(denoted, in the following, $l_2$) of the second peak. In addition, we
show that the location of the second peak in the quintessence case and
in the cosmological constant case can be easily
distinguished. Following Ref.~\cite{KB}, we display the contour plots
of $l_2$ in the plane $(\Omega_\MAT, h)$ for the Ratra-Peebles and
SUGRA tracking potentials.
\par
The article is organized as follows. In section~\ref{sec_bg_evol}, we
give a description of the background evolution in terms of two
physical quantities: the equation of state parameter $\omega_\QUINT$
and the sound velocity $\vson_\QUINT$. In section~\ref{sec_cosm_pert},
we study the cosmological perturbations for the quintessence field. In
section~\ref{sec_pred_cl}, we present the results of full numerical
calculations with the help of a Boltzmann code developped by one of us 
(A.R.) for the CMB anisotropies and power spectra in the case of
the Ratra-Peebles and SUGRA potentials. Then, detailed comparisons
with the recent BOOMERanG~\cite{B98a,B98b} and
MAXIMA-1~\cite{MAX1a,MAX1b} data are performed. We end with our main
conclusions in section~\ref{sec_conc}.

\section{The background evolution}
\label{sec_bg_evol}

We suppose that the Universe can be described by a
Friedman-Lema\^\i{}tre-Robertson-Walker metric the spacelike sections
of which are flat
\begin{equation}
\label{defmetric}
\ddd s^2 = a^2(\eta) (-\ddd \eta ^2 + \delta_{ij} \ddd x^i \ddd x^j).
\end{equation}
In this equation, $\eta$ is the conformal time related to the cosmic
time by $a(\eta) \ddd \eta \equiv \ddd t$. The matter content is
as follows. The Universe is filled with a mixture of five fluids:
photons ($\gamma$), neutrinos ($\nu$), baryons ($\BAR$), cold dark
matter ($\CDM$) and a scalar field $Q$ named quintessence. The stress
energy tensor of each of these species is the one of a perfect fluid,
$T_{\mu \nu} = (p + \rho) u_\mu u_\nu + p g_{\mu \nu}$, where
$u_\mu$ is the 4-velocity of the fluid. The energy density and the
pressure of the scalar field are given by $\rho_\QUINT = \frac{1}{2}
(Q'/a)^2 + V(Q)$ and $p_\QUINT = \frac{1}{2} (Q'/a)^2 - V(Q)$, 
where $V(Q)$ is the potential of quintessence whose shape will be very
important in what follows. Each fluid is also characterized by its
equation of state $p_{\rm i} \equiv \omega _{\rm i}\rho_{\rm i}$ 
where ${\rm i} = \gamma, \nu, \BAR, \CDM$ or $\QUINT$. We have
$\omega_\gamma = \omega_\nu = 1 / 3$ and $\omega_\BAR =
\omega_\CDM = 0$. The case of $\omega_\QUINT$ is more complicated
since this is a time-dependent function such that $-1 \le
\omega_\QUINT \le +1$. Its expression reads $\omega_\QUINT = 1 - 2
V(Q) / \rho_\QUINT$. The fact that $\omega_\QUINT$ is a time-dependent
function directly comes from the fact that, for a scalar field, the
sound velocity defined as~\cite{MS}
\begin{equation}
\label{defsoundv}
\vson_\QUINT \equiv \frac{p_\QUINT'}{\rho_\QUINT'}
 = 1 + \frac{4 a^2}{3 \Hconf Q'} \frac{\ddd V(Q)}{\ddd Q} 
 = - \frac{1}{3} \left(2 \frac{Q''}{\Hconf Q'} + 1\right),
\end{equation}
is not equal to the equation of state parameter $\omega_\QUINT$. As a
consequence $\omega_\QUINT$ has to change in time as revealed by the
following equation
\begin{equation}
\label{timeomega}
\omega_\QUINT ' = - 3 \Hconf (1 + \omega_\QUINT) 
                             (\vson_\QUINT - \omega_\QUINT) ,
\end{equation}
unless $\omega_\QUINT = -1$.
\par
The evolution of the Universe can be calculated with the help of the
Friedman and conservation equations
\begin{eqnarray}
\label{fried}
\frac{1}{a^2} \Hconf^2 & = & \frac{8 \pi}{m_\PL^2} \sum_i \rho_{\rm i}, \\
\label{conser}
\rho_{\rm i}'          & = & -3 \Hconf (1 + \omega_{\rm i}) \rho_{\rm i}, 
\quad {\rm i} = \gamma, \nu, \BAR, \CDM \mbox{ or } \QUINT,
\end{eqnarray}
where $m_\PL$ is the Planck mass and $\Hconf \equiv a' / a$ is related
to the Hubble constant by the equation $H = \Hconf / a$. The equations
of conservation simply express the fact that the energy is conserved
for each species which do not interact. The equation of conservation
of the quintessence field can also be written as the Klein-Gordon
equation
\begin{equation}
\label{KG}
Q'' + 2 \Hconf Q' + a^2 \frac{\ddd V}{\ddd Q} = 0 .
\end{equation}
We now need to give the last piece of information necessary to have a
complete description of the system, \IE{} the shape of the potential
$V(Q)$. In order to be an interesting theory and to represent an
improvement over the current situation, quintessence has to address
the following four problems: the fine-tuning problem, the coincidence
problem, the equation of state problem and the model building
problem. The fine-tuning problem amounts to understanding whether one
can have $\Omega_\QUINT \simeq 0.7$ with the free parameters of the
potential taking ``natural'' values, \IE{} close to the energy scale
of the theory under consideration. The coincidence problem is the
question of the initial conditions: does the final value of
$\rho_\QUINT$ strongly depend on the chosen initial values of $Q$ and
$Q'$?  The equation of state problem is the question of the value of
$\omega_\QUINT$. In order to be compatible with observational data, it
should be such that $-1 < \omega_\QUINT < 0$. According to recent
papers, even more stringent restrictions can be put, namely
$-1<\omega_\QUINT < -0.6$~\cite{WCOS} or even $-1 < \omega_\QUINT <
-0.8$~\cite{Efst}. In particular, this already rules out a network of
cosmic strings since the corresponding fluid has an equation of state
parameter equal to $- 1 / 3$. Finally, the model building problem
consists in justifying the shape of the potential from the high energy
physics point of view. Many different shapes of potential which allow,
at least partially, to solve these problems have been investigated in
the litterature and Table~\ref{potlist} summarizes these
proposals.
\begin{table}
\begin{tabular}{c c}
Potential  & References \\ \hline
\\ 
$\Lambda^{4 + \alpha } / Q^\alpha$               & \cite{RP} \\ \\
$\Lambda^4 e^{- \lambda Q}$                      & \cite{RP,FJ1} \\ \\ 
$(\Lambda^{4 + \alpha } / Q^\alpha) e^{\frac{\kappa}{2} Q^2}$
                                                 & \cite{BM2,BM1} \\ \\
$\Lambda^4 (\cosh \lambda Q - 1)^p$              & \cite{SW} \\ \\ 
$\Lambda^4 (e^{\alpha \kappa Q} + e^{\beta \kappa Q})$
                                                 & \cite{BCN} \\ \\ 
$\Lambda^4 e^{- \lambda Q} (1 + A \sin \nu Q)$   & \cite{DKS} \\ \\ 
$\Lambda^4 [(Q - B)^\alpha + A] e^{- \lambda Q}$ & \cite{AS} \\ \\
$\Lambda^4 [1 + \cos (Q / f)]$                   & \cite{Kim} \\
\end{tabular}
\caption{Quintessence potentials that have been used in the 
litterature.}
\label{potlist}
\end{table}
In particular, the first possibility has been studied thoroughly in
the past years. In this article, we will mainly concentrate on the
Ratra-Peebles potential~\cite{RP} and the SUGRA tracking
potential~\cite{BM2,BM1}.
\par
Let us briefly see how the four questions evoked previously can be
addressed with these potentials.

\subsection{The fine-tuning problem}

Let us start with the fine-tuning problem which is clearly a delicate
question. This problem is crucial~\cite{Wein} for the cosmological
constant. Indeed, from very simple high energy physics considerations,
one typically expects $\rho_\Lambda \simeq m_\PL^4 \simeq 10^{76}
\UUNIT{GeV}{4}$ whereas one measures $\rho_\Lambda \simeq
\Omega_\Lambda \rho_\CRIT \simeq 10^{-47} \UUNIT{GeV}{4}$ since the
critical energy density is $\rho_\CRIT \simeq 8.1 h^2 \times 10^{-47}
\UUNIT{GeV}{4}$. Do we gain something in the case of quintessence?
This question is controversial. For example in Ref.~\cite{BBS}, the
authors clearly answer no and write ``Two proposals to explain these
observations are a non-vanishing cosmological constant or a very
slowly rolling scalar field, often dubbed {\it quintessence}. Both
proposals, however, are plagued with formidable fine tuning
problems.'' However, one should look more carefully at this point. To
illustrate this issue, let us consider the general argument given
against quintessence. If we consider the potential $V(Q) = (m^2 / 2)
Q^2$ then the mass of such a field, which is also the only free
parameter of the potential, should be $m = \sqrt{2 \Omega_\QUINT
\rho_\CRIT} / m_\PL \simeq 10^{-33} \UUNIT{eV}{}$, a very tiny mass
indeed.  Justifying such a value for the free parameter $m$ is
probably the same problem as justifying a very low value for
$\rho_\Lambda$. However, such a model has never been advocated for the
quintessence field. As already mentioned above, one typically
considers models such that $V(Q) = \Lambda^{4 + \alpha } /
Q^\alpha$. This changes the argument. Now, the free parameter of the
theory is $\Lambda$. In order to have $\rho_\QUINT = \Omega_\QUINT
\rho_\CRIT$ today, one has $\Lambda \simeq 10^{11} \UUNIT{GeV}{}$, for
$\alpha = 11$. This time, the free parameter of the theory has a value
comparable to the natural scales of high energy physics. Therefore,
something has been gained and it seems unfair not to emphasize this
point. On the other hand the mass of the field is given by $m = \alpha
(\alpha + 1) \Omega_\QUINT \rho_\CRIT /m_\PL^2 \simeq 10^{-33}
\UUNIT{eV}{}$ but this number should be interpreted completely
differently. Here the mass $m$ is just a ``by-product'' and its value
is naturally very small without any artificial fine-tuning of
$\Lambda$. Of course the very small value of the mass implies that the
quintessence field is almost completely decoupled from the other
matter fields. This renders the model building issue even more acute.

\subsection{The coincidence problem}

The coincidence problem as formulated in the introduction, \IE{} the
dependence upon the initial conditions, is solved because the
Klein-Gordon equation possesses an attractor. In order to prove this
property, we have to rely either on numerical calculations or on
approximate methods. All the plots and numerical estimates displayed
in this article will be made with the help of numerical
calculations. However, it is always useful to understand the tracking
property by means of analytical methods and we now turn to this
question. It is convenient, for analytical calculations, to consider
that there is in fact only one ``background'' fluid with a time
dependent equation of state such that $\omega_\BG = 1 / 3$ during the
radiation dominated epoch and $\omega_\BG = 0$ during the matter
dominated era. In addition to the background fluid, we assume that
there also exists the quintessence scalar field field. Following the
treatment of Ratra and Peebles \cite{RP}, it will be considered that
this scalar field is a test field. This is a good approximation since
this field must be sub-dominant in particular during Big Bang
Nucleosynthesis (BBN) in order not to modify the behaviour of the
scale factor and, as a consequence, not to spoil the success of
BBN. This means that the behaviour of the scale factor is essentially
determined by the background fluid and that $\sum_i \rho_{\rm i}
\simeq \rho_\BG$. This hypothesis breaks down at very small redshift when
quintessence starts dominating the matter content of the
Universe. Since quintessence is only a test field which does not
interact with the background fluid, the scale factor and the quantity
$\Hconf$ can be written as
\begin{equation}
\label{scalef}
a(\eta) \propto \eta ^{2 / (1 + 3 \omega_\BG)}
\quad,\quad
\Hconf(\eta) = \frac{2}{(1 + 3\omega_\BG) \eta}.
\end{equation}
For the sake of illustration, let us now consider the radiation
dominated era where $\omega_\BG = 1 / 3$. Under the previous
assumptions, the Klein-Gordon equation has a particular solution given
by
\begin{equation}
\label{partsol}
Q_{\rm p} = Q_0 \eta ^{4 / (\alpha + 2)},
\end{equation}
where $Q_0$ is a constant which depends on the free parameters of the
potential, \IE{} $\Lambda$ and $\alpha$. The tracking behaviour is
revealed by the behaviour of small perturbations around $Q_{\rm
p}$. Let us introduce the new time $\tau$ defined by $\eta \equiv
e^\tau$ and define $u$ and $p$ by $Q = Q_{\rm p} u$ and $p = \ddd
u/\ddd \tau $. The Klein-Gordon equation, viewed as a dynamical system
in the plane $(p, u)$, possesses a critical point $(0, 1)$ and small
perturbations around this point $\delta u, \delta p$ obey the
following equation
\begin{eqnarray}
\label{KGmatrix}
\frac{\ddd}{\ddd \tau} \pmatrix{\delta p \cr
                                \delta u}
 =  \pmatrix{  - \frac{\alpha + 10}{\alpha + 2}
             & - \frac{4 (\alpha + 6)}{\alpha + 2} \cr 
               1 
             & 0}
    \pmatrix{\delta p \cr 
             \delta u}.  
\end{eqnarray}
Solutions to the equation $\mbox{det}(M - \lambda I) = 0$, where $M$
is the matrix defined above, are given by
\begin{equation}
\label{sol}
\lambda_\pm = - \frac{\alpha + 10}{2 (\alpha + 2)} 
              \pm \frac{i}{2 (\alpha + 2)} 
                  \sqrt{15 \alpha^2 + 108 \alpha + 92}.
\end{equation}
The real part of $\lambda_\pm$ is always negative and the critical
point is a spiral point. Therefore, every solution will tend to $Q =
Q_{\rm p}$ after an intermediate regime: $Q = Q_{\rm p}$ is an
attractor and no fine-tuning of the initial conditions is required.
\par
Before reaching the attractor, the quintessence field undergoes
different regimes that we are now going to describe. These regimes are
in fact characterized by two physical quantities already introduced
previously: the equation of state parameter $\omega_\QUINT$ and the
sound velocity $\vson_\QUINT$. We study the case of an ``overshoot'',
in the terminology of Ref.~\cite{SWZ}, since this corresponds to
initial conditions that are physically more relevant (in particular
this includes the case of equipartition, \IE{} $\rho_\QUINT
\simeq 10^{-4} \rho_\BG$ initially). We also assume that 
the background is radiation dominated $\omega_\BG = 1 / 3$.
\par
\begin{figure}
\centerline{\psfig{file=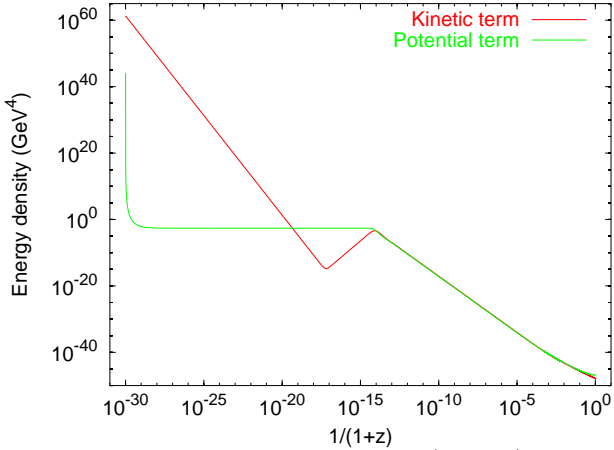,width=3.5in,angle=-90}}
\caption{Evolution of the kinetic energy (red line) and of
the potential energy $V(Q)$ (green line) from $z_{\rm i} =
10^{30}$ to $z=0$.}
\label{KV}
\end{figure}

Initially, the kinetic energy dominates the potential energy, \IE{}
$Q'^2 / (2 a^2) \gg V(Q)$. This means that the energy density
redshifts as $\rho_\QUINT \propto 1 / a^6$ and that the equation of
state parameter is $\omega_\QUINT = 1$. As a consequence, due to the
constancy of $\omega_\QUINT$ and Eq.~(\ref{timeomega}) (and also
$\omega_\QUINT \neq -1$), we have $\vson_\QUINT = 1$ as well. The
scalar field itself evolves like
\begin{equation}
\label{fieldKdom}
Q = Q_{\rm f} - \frac{A}{a},
\end{equation}
where $Q_{\rm f}$ and $A$ are constant. These constants are such that
the term $A / a$ becomes rapidly small in comparison with the frozen
value $Q_{\rm f}$ and we have the amusing situation that the field can
be (almost) considered as frozen even if the kinetic energy still
dominates. This is illustrated in Fig.~\ref{KV}.

As a consequence, during this regime the potential energy is also
almost constant except at the very beginning. Using the definition of
$\omega_\QUINT$ and $\vson_\QUINT$, see Eq.~(\ref{defsoundv}), we
deduce that, during the kinetic regime, we have
\begin{equation}
\label{evolwc}
\omega_\QUINT - 1 \propto a^6, \quad \vson_\QUINT - 1  \propto a^5.
\end{equation}
The fact that, in the parametrization adopted here, the scale factor
is very small during the kinetic regime explains that there is no
contradiction between these equations and the values of
$\omega_\QUINT$ and $\vson_\QUINT$ deduced above.
\par
Since the kinetic energy decreases while the potential energy is
almost constant, the kinetic regime cannot last forever. When the
potential energy becomes larger than the kinetic one, the equation of
state parameter suddenly jumps from $+1$ to $-1$ while the sound
velocity still remains equal to $+1$ since Eq. (\ref{timeomega}) does
not imply a change of this quantity in the case $\omega_\QUINT = -
1$. The fact that the equation of state parameter changes before the
sound velocity is explained by Eq.~(\ref{evolwc}). We call this regime
the transition regime. During this regime, the kinetic energy still
redshifts as $1 / a^6$ and $V(Q)$ is approximately constant but of
course now $\rho_\QUINT \simeq V(Q)$.
\par
\begin{figure}
\centerline{\psfig{file=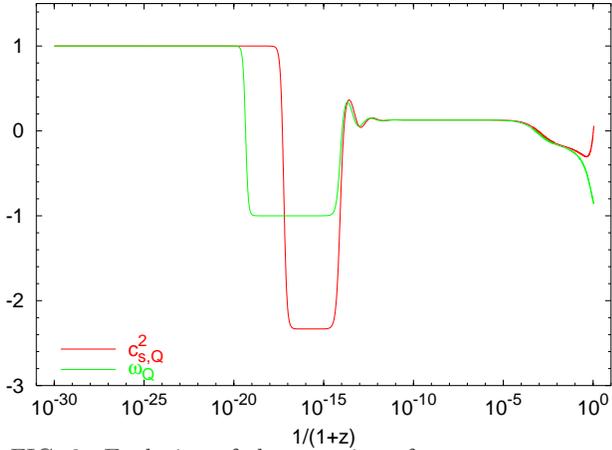,width=3.5in,angle=-90}}
\caption{Evolution of the equation of state parameter $\omega_\QUINT$ 
(green line) and of the sound velocity $\vson_\QUINT$ 
(red line) from $z_{\rm i} = 10^{30}$ to $z=0$.}
\label{omv}
\end{figure}

Due to the second of Eq.~(\ref{evolwc}), the sound velocity has also
to change at some later time. This implies that the quintessence field
can no longer behave according to Eq.~(\ref{fieldKdom}). This is the
starting point of the potential regime. In order to study the
behaviour of the system in this regime, we need to find an expression
for the second derivative of the potential. Differentiating once the
definition of the sound velocity, Eq.~(\ref{defsoundv}), we arrive at
\begin{eqnarray}
\label{genetrack}
\frac{\ddd ^2 V(Q)}{\ddd Q^2}
 & = & \frac{3}{2} H^2 \biggl\{\frac{1}{\Hconf} \vson_\QUINT ' \nonumber \\
 & + & (\vson_\QUINT - 1) \biggl[  \frac{\Hconf'}{\Hconf^2}
                                 - \frac{1}{2} (3 \vson_\QUINT + 5)
                          \biggr] 
                       \biggr\}.
\end{eqnarray}
No approximation has been made in the derivation of this
relation. This formula generalizes Eq.~(3) of Ref.~\cite{ZWS}. This
formula will turn out to be very useful when we study the
perturbations in the next section. With the scale factor given by
Eqns.~(\ref{scalef}), this relation can be re-written as
\begin{equation}
\label{potregime}
\frac{2}{3 H^2} \frac{\ddd ^2 V(Q)}{\ddd Q^2}
 =   \frac{1}{\Hconf} \vson_\QUINT '
   - 3 (\vson_\QUINT - 1) (\omega_\BG + \vson_\QUINT + 2).
\end{equation}
In the regime we are interested in, the r.h.s of the previous formula
is small. The only way to satisfy this relation is to ensure that the
sound velocity changes to the constant $\vson_\QUINT = - 2 -
\omega_\BG$. This gives $\vson_\QUINT = - 7 / 3$ for the radiation
dominated era. This evolution is displayed in Fig.~\ref{omv}.

The fact that the sound velocity is a constant implies that the factor
$(4 a^2) / (3 \Hconf Q') \ddd V(Q)/ \ddd Q$ is also a
constant. Therefore, the behaviour of the quintessence field is now
given by
\begin{equation}
\label{qpot}
Q = Q_{\rm f} + B a^4,
\end{equation}
which implies that the kinetic energy redshifts as $a^4$.
\par
Again this regime cannot last forever since the kinetic energy
increases while the potential energy still remains constant. At some
later time, both contributions become equal and $\omega_\QUINT$ and
$\vson_\QUINT$ have to change once more. This is the end of the
potential regime and the beginning of the tracking regime which has
already been described above. The quantities $p_\QUINT$,
$\rho_\QUINT$, $V$ and the kinetic energy reach a fixed ratio such
that
\begin{equation}
\label{tracking}
\omega_\QUINT = \vson_\QUINT = - \frac{2 - \alpha\omega_B}{2 + \alpha}.
\end{equation}
The definitions of the different regimes and the corresponding
evolutions of the physically relevant quantities are summed up in 
Table~\ref{summary}.

\begin{table}
\begin{tabular}{c c c c c c} 
Regime & $Q'^2/(2a^2)$ & $V(Q)$ & $Q$ & $\omega_\QUINT$ &
$\vson_\QUINT$ \\ \hline
\\
Kinetic    & $1 / a^6$ &  $a^0$ & $Q_{\rm f} - A / a$ & $1$  & $1$ \\ 
Transition & $1 / a^6$ &  $a^0$ & $Q_{\rm f} - A / a$ & $-1$ & $1$ \\ 
Potential  & $a^4$     &  $a^0$ & $Q_{\rm f} + B a^4$ & $-1$ & $- 7 / 3$ \\ 
Tracking   & $a^{-3(1+\omega_\QUINT)}$ & $a^{-3(1+\omega_\QUINT)}$ &  
             $a^{4 / (\alpha + 2)}$ & 
             $\frac{\alpha \omega_\BG - 2}{\alpha + 2}$ & 
             $\omega_\QUINT$ 
\end{tabular}
\caption{Summary of the different regimes described in the text.}
\label{summary}
\end{table}

\subsection{The equation of state parameter problem}

The third question evoked previously was the question of the value of
the parameter $\omega_\QUINT$ today. As already mentioned, this is an
important issue since constraints on this quantity are already
available. This problem is also solved by quintessence in the sense
that we always have $- 1 < \omega_\QUINT < 0$. Here, however, it is
relevant to distinguish between the Ratra-Peebles potential and the
SUGRA potential. With the first potential, it seems difficult to reach
sufficiently small value of $\omega_\QUINT$. On the other hand, this
is automatically achieved in the second case. The reason for this is
the presence of the factor $\exp(\kappa Q^2 / 2)$ in the potential, a
generic feature of SUGRA-based potentials, which drives
$\omega_\QUINT$ towards $-1$. For $\alpha = 11$ and $\Omega_\CDM
\simeq 0.3$, the prediction is $\omega_\QUINT \simeq -0.82$ a value in
agreement with the current data~\cite{BM2,BM1}.

\subsection{The model building problem}

{}From the particle physics point of view, one would like to justify
the existence of the quintessence field and the shapes of the (so far)
phenomenological potentials. Several attempts have already been made
in the framework of supersymmetric field theory. In particular, it was
shown by Bin\'etruy~\cite{Bine} that the Ratra-Peebles potential can
be recovered in the context of global SUSY. However, as already
mentioned, SUGRA corrections must be taken into account and this
implies that the corresponding potential can be of the type of the
SUGRA tracking potential displayed in Eq.~(\ref{citepot}) which leads
to a better agreement with the available data.
\par
Nevertheless, it should be clear that considerable problems remain to
be addressed in order to reach a satisfactory
situation~\cite{Carroll,KL}. Maybe the most crucial question is the
problem of SUSY breaking. SUSY must certainly be broken but the models
evoked previously do not take into account this basic fact. This could
have dramatic consequences and modify the shape of the potential which
is so important in order to solve the three previous problems.

\section{The cosmological perturbations}
\label{sec_cosm_pert}

We now turn to the study of the cosmological perturbations. A 
detailed study has already been performed by Ratra and Peebles in
Ref.~\cite{RP} but only for the tracking regime. Cosmological
perturbations in a fluid with a constant negative equation of state
parameter have been investigated in Ref.~\cite{FM}. In this article,
we study the cosmological perturbations (in the long wavelength
approximation) in all the regimes previously described and point out
some additional properties. The evolution of the cosmological
perturbations mainly depends on the equation of state parameter and
the sound velocity. We have shown in the previous section that they
can be considered as constant in each regime. This will simplify the
analysis a lot.
\par
The fate of the perturbations depends on the initial conditions. It
has been noticed for the first time in Ref.~\cite{CDS} that ``the
observable fluctuation spectrum is insensitive to a broad range of
initial conditions, including the case in which the amplitudes of
$\delta Q$, $\delta Q'$ are set by inflation''. In that paper, the
authors choose $\delta Q = \delta Q' = 0$ initially (in the
synchronous gauge). We demonstrate, in this section, that the
insensitivity of the spectrum described in Ref.~\cite{CDS} has an
origin similar to the insensitivity of the background properties with
respect to the initial conditions $Q$ and $Q'$, namely the presence of
an attractor for the perturbed quantities. We prove that during all
the four regimes undergone by the quintessence field, the attractor is
characterized by a ``spiral fixed point'' as it is the case for the
background.

\subsection{General framework}

Without loss of generality, the perturbed line element can be written
in the synchronous gauge. In this class of coordinates systems, scalar
perturbations are completely described by two arbitrary functions. The
spatial dependence of the perturbations is given by $X(x^i)$ which is
the eigenfunction of the Laplace operator on the flat spacelike
hypersurfaces. There exists two ways to construct a two rank tensor
from a scalar function~: either by multiplying it by the spatial
background flat metric $\delta_{ij}$ or by differentiating it
twice. The two arbitrary functions mentioned above are simply the
coefficients of these two terms in a Fourier expansion. Therefore, the
perturbed metric can be expressed as~\cite{Gri}
\begin{eqnarray}
\label{metricsg}
\ddd s^2 = a^2(\eta) \biggl\{ & - &\ddd \eta^2
                        + \biggl[  \biggl(1 + h(\eta) X \biggr) \delta_{ij}
\nonumber \\ 
                              & + & h_l(\eta) \frac{1}{k^2} X_{,i,j}
                          \biggr] \ddd x^i \ddd x^j
                \biggr\}.
\end{eqnarray}
In this equation, the dimensionless quantity ${\bf k}$ is the comoving
wavevector related to the physical wavevector ${\bf k}^{\rm phys}$
through the relation ${\bf k}^{\rm phys} \equiv {\bf k} / a(\eta)$. As
a consequence of Einstein equations, perturbations in the metric are
coupled to perturbations in the different matter components. We choose
to write the perturbed stress-energy tensor according to~\cite{Gri}
\begin{eqnarray}
\label{stresener}
T^0{}_0 & = & - \frac{\epsilon_1}{a^2} X
, \quad 
T^0{}_i   =     \frac{\xi'}{a^2} X_{,i}
, \quad 
T^i{}_0   =   - \frac{\xi'}{a^2}X ^{,i}, \\
T^i{}_j & = &   \frac{p_1}{a^2} X \delta^i{}_j,
\end{eqnarray}
where we have assumed that the longitudinal pressure $p_l$ vanishes
for each component. As for the background, one considers that the
Universe is filled with two fluids: the background fluid, an
hydrodynamical perfect fluid which is either radiation or dust (again,
the corresponding quantities will carry the index $\BG$) and a scalar
field $Q$ describing the quintessence field (in this case the
corresponding quantities will carry the index $\QUINT$). The perturbed
Einstein equations which govern the evolution of the quantities $h$
and $h_l$ are given by:
\begin{eqnarray}
\label{e1}
3 \Hconf h' + k^2 h - \Hconf h_l'
 & = & \kappa \epsilon_{1\BG} + \kappa \epsilon_{1\QUINT}, \\
\label{e2}
h' & = & \kappa \xi'_\BG + \kappa \xi'_\QUINT, \\
\label{e3}
- h'' - 2 \Hconf h' & = & \kappa p_{1\BG} + \kappa p_{1\QUINT}, \\
\label{e4}
h_l' + 2 \Hconf h_l' - k^2 h & = & 0.
\end{eqnarray}
Finally, it turns out to be more convenient to work with the density
contrast $\delta$ and the velocity divergence $\theta$ defined by the
equations:
\begin{equation}
\label{defss}
\delta \equiv \frac{\epsilon_1}{a^2 \epsilon_0}
\quad,\quad 
\xi' \equiv - \frac{a^3 \epsilon_0}{k^2} (1 + \omega) \theta .
\end{equation}
In the following, we study analytically the time evolution of the
density contrast for the background fluid and for quintessence in the
long wavelength limit.

\subsection{The background fluid}

The equations satisfied by the background density contrast and
divergence can be obtained either from combinations of the Einstein
equations (\ref{e1}-\ref{e4}) or, more directly, from the
conservation of the perturbed background fluid stress-energy tensor
(since the background fluid and quintessence only interact
gravitationally). They read:
\begin{eqnarray}
\label{conserv1}
\delta_\BG' + a (1 + \omega_\BG) \theta_\BG
            + \frac{1 + \omega_\BG}{2} (3 h' - h_l') & = & 0, \\
\label{conserv2}
\theta_\BG' + (2 - 3 \omega_\BG) \Hconf \theta_\BG 
            - \frac{k^2 \vson_B}{(1 + \omega_\BG) a} \delta_\BG & = & 0 .
\end{eqnarray}
These two equations are equivalent to Eqns.~(7.15) and~(7.16) of
Ref.~\cite{RP}. From them, we can derive the relation
\begin{equation}
\label{trace''}
3h'' - h_l''
 =   \frac{- 2}{1 + \omega_\BG} \delta_\BG ''
   + 2 (1 - 3 \omega_\BG) a' \theta_\BG 
   - \frac{2 k^2 \vson_B}{1 + \omega_\BG} \delta_\BG ,
\end{equation}
where we have assumed that $\omega_\BG$ is a constant. On the other
hand, from the Einstein equations we get
\begin{eqnarray}
\label{trace''and'}
& & - (3 h'' - h_l'') - \Hconf (3 h' - h_l') \nonumber \\
& & = 3 \Hconf^2 \biggl[  (1 + 3 c_{\ell \BG}^2) \Omega_\BG \delta_\BG
                        + (1 + 3 c_{\ell \QUINT}^2) \Omega_\QUINT \delta_\QUINT
                 \biggr],
\end{eqnarray}
where $c_{\ell \QUINT}^2 \equiv p_{1\QUINT} / \epsilon_{1\QUINT}$
which needs not to coincide with the definition of $\vson_\QUINT$. In
order to derive the formula satisfied by the density contrast of the
background fluid in the long wavelength limit, we neglect the term
proportional to $k^2$ in Eq.~(\ref{trace''}) and we use the fact that
$\Omega_\QUINT \ll \Omega_\BG$. Then, straightforward manipulations
lead to
\begin{eqnarray}
\label{contraB} 
\delta_\BG'' + \Hconf \delta_\BG'
 & - & \frac{3}{2}\Hconf^2 (1 + 3 \omega_\BG) (1 + \omega_\BG)\delta_\BG
\nonumber \\
 & = & - 3 \Hconf \omega_\BG (1 + \omega_\BG) a \theta_\BG ,
\end{eqnarray}
where we used the fact that $c_{\ell \BG}^2 = \omega_\BG$ for an
hydrodynamical fluid. This equation shows that the evolution of the
background density contrast is essentially unaffected by the presence
of quintessence. This is of course an expected result since we have
assumed $\Omega_\QUINT \ll \Omega_\BG$. The general solution to
Eq.~(\ref{contraB}) can be easily found and reads
\begin{eqnarray}
\label{solcontraB}
 &   & \delta_\BG(\eta) =   A_1 \biggl(\frac{a}{a_0} \biggr)^{x_+}
                          + A_2 \biggl(\frac{a}{a_0} \biggr)^{x_-} 
\nonumber \\ 
 & + & \frac{\omega_\BG (1 + \omega_\BG) 
             (1 + 3 \omega_\BG) a_0 \theta_{\BG 0} \eta_0} 
            {(1 - \omega_\BG) (1 + 6 \omega_\BG)}
       \biggl(\frac{a}{a_0} \biggr)^{(9 \omega_\BG - 1) / 2},
\end{eqnarray}
where we have defined
\begin{eqnarray}
\label{racine}
x_\pm & \equiv & - \frac{(1 - 3 \omega_\BG)}{4} \nonumber \\
      &        & \pm \frac{1}{4} 
                     \sqrt{  (1 - 3 \omega_\BG)^2
                           + 24 (1 + \omega_\BG ) (1 - 3 \omega_\BG )}.
\end{eqnarray}
The results for the radiation dominated and matter dominated epochs
are summarized in Table~\ref{last_table}.
\begin{table}
\begin{tabular}{c c c c}
$\omega_\BG $ & $x_-$     & $x_+$ & $(9 \omega_\BG - 1) / 2$ \\ \hline
$1 / 3$       & $-2$      & $2$   & $1$ \\
$0$           & $- 3 / 2$ & $1$   & $- 1 / 2$ 
\end{tabular}
\caption{Time dependence of the background fluid density contrast
during radiation and matter dominated era}
\label{last_table}
\end{table}
These results are consistent with those obtained in
Ref.~\cite{Pee}. In particular, it can be shown that the branch
$\delta_\BG \propto a^{x_-}$ corresponds in fact to a residual gauge
mode, \IE{} there exists a synchronous system of coordinates such that
this mode can be removed and therefore must not be considered as a
physical mode.

\subsection{Quintessential perturbations}

We now describe how the long wavelength quintessential perturbations
evolve with time. A similar study has already been performed by Ratra
and Peebles but only on the tracking solution. We give here a complete
description of the evolution of the quintessence density contrast in
the four regimes defined in the previous section. In addition, we
prove that there exists an attractor for the perturbations as it is
the case for the background solution. As a consequence, the final
value of the density contrast is always the same whatever the initial
conditions are.

In order to obtain the fundamental equations to be solved, we can
proceed as for the background fluid. However, it is important to
notice that the link between the perturbed energy density and the
perturbed pression, which is just a constant for the background fluid,
is more complicated in the case of quintessence. In general, we can
write $p_{1\QUINT} = \vson_\QUINT \epsilon_{1\QUINT} + a^2 \tau \delta
S$ where the second term represents entropy perturbations. In the
synchronous gauge, we obtain
\begin{equation}
\label{velosf}
p_{1\QUINT} =   \vson_\QUINT \epsilon_{1\QUINT}
              + (1 - \vson_\QUINT) \frac{1}{\kappa} (h_l'' + \Hconf h_l').
\end{equation}
\par
\begin{figure}
\centerline{\psfig{file=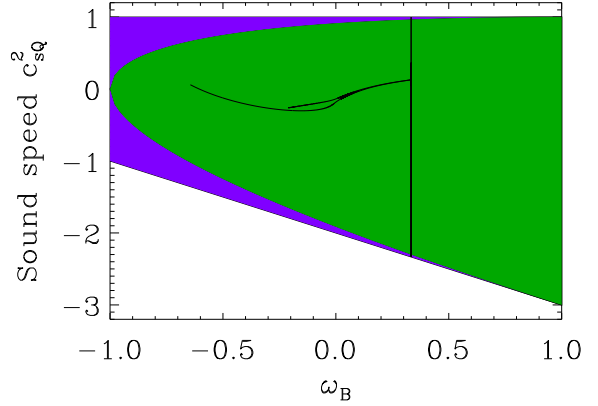,width=3.5in,angle=0}}
\caption{Stability region for the quintessential
perturbations. The green (lightgrey) and the purple regions (darkgrey)
constitute the stable region. The green region is the region where the
square root is an imaginary number. Two trajectories of the system for
the tracking potential (shortest line) and the SUGRA tracking
potential (longest line) are also displayed for the value $\alpha =
11$.}
\label{toto}
\end{figure}

We can now establish the equations satisfied by the quintessence
density contrast and divergence. The conservation of the perturbed
stress-energy tensor leads to
\begin{eqnarray}
\label{cons0sf}
&   & \delta_\QUINT' + 3 \alpha (\vson_\QUINT - \omega_\QUINT) \delta_\QUINT
                     + a (1 + \omega_\QUINT) \theta_\QUINT 
\nonumber \\ 
& + & \frac{1}{2} (1 + \omega_\QUINT) (3 h' - h_l')
 = \frac{\Hconf^{-1}}{\Omega_\QUINT} 
   (\vson_\QUINT - 1) (h_l'' + \Hconf h_l'), \\
\label{consisf}
&   & \theta_\QUINT' + (2 - 3 \omega_\QUINT) \alpha \theta_\QUINT 
                     - \frac{k^2 \vson_\QUINT}{(1 + \omega_\QUINT) a}
                       \delta_\QUINT  \nonumber \\
& = & - \frac{\omega_\QUINT'}{1 + \omega_\QUINT} \theta_\QUINT
      + \frac{(1 - \vson_\QUINT) k^2}
             {3 a \Hconf^2 (1 + \omega_\QUINT) \Omega_\QUINT}
        (h_l'' + \Hconf h_l').
\end{eqnarray}
In these two equations, no approximation have been made: they are
valid for any wavenumber, any equation of state parameter and sound
velocity. In practice, it turns out to be more convenient to use the
perturbed Klein-Gordon equation to analyse the problem. This can be
obtained directly from the first of the two previous equations if one
notices that the quantities describing the perturbed scalar field
stress-energy tensor can be expressed in terms of the perturbed scalar
field $\delta Q(\eta, {\bf x})$ according to
\begin{eqnarray}
\label{sf1}
\epsilon_{1\QUINT}
 & = & Q' \delta Q' + a^2 \delta Q \frac{\ddd V(Q)}{\ddd Q} , \\
\label{sf2}
\xi_{\QUINT}' & = & - Q' \delta Q , \\
\label{sf3}
p_{1\QUINT} & = & Q' \delta Q' - a^2 \delta Q \frac{\ddd V(Q)}{\ddd Q}.
\end{eqnarray}
Inserting the corresponding expression for the density contrast and
the divergence in Eq.~(\ref{cons0sf}), we get
\begin{equation}
\label{KGpert}
\delta Q'' + 2 \Hconf \delta Q' 
           + \biggl[k^2+a^2 \frac{\ddd ^2 V(Q)}{\ddd Q^2} \biggr] \delta Q 
           + \frac{Q'}{2} (3 h' - h_l') = 0.
\end{equation}
This is similar to Eq.~(7.20) of Ref.~\cite{RP}. One can check that
Eq.~(\ref{consisf}) is automatically verified since it is equivalent
to the unperturbed Klein-Gordon equation (times an unimportant
factor). Using Eq.~(\ref{conserv1}) to express the factor $3 h' -
h_l'$ and neglecting the $k^2$ term, we arrive at
\begin{equation}
\label{KGlwave}
\delta Q'' + 2 \Hconf \delta Q'
           + a^2 \frac{\ddd ^2 V(Q)}{\ddd Q^2} \delta Q
 = Q' a \theta_\BG + \frac{Q'}{1 + \omega_\BG} \delta_\BG'.
\end{equation}
We are now going to analyse this equation in detail. We now need to
utilize the general expression for the second derivative of the
potential, Eq.~(\ref{genetrack}). On the tracking solution, we have
$\omega_\QUINT = \vson_\QUINT$ and $\omega_\QUINT = (- 2 + \alpha
\omega_\BG) / (2 + \alpha)$ and this equation reduces to
\begin{equation}
\label{trackStein}
\frac{\ddd^2 V(Q)}{\ddd Q^2} = \frac{9}{2} H^2 \frac{\alpha + 1}{\alpha}
                               (1 - \omega_\QUINT^2).
\end{equation}
For our purpose, as proven in the previous section, it is sufficient
to consider a regime where $\vson_\QUINT$ is constant and where the
scalar field is a test field. Under these conditions, we obtain
\begin{equation}
\label{4regimes}
\frac{\ddd ^2 V(Q)}{\ddd Q^2} = \frac{3}{4} H^2 (1 - \vson_\QUINT)
                                (6 + 3 \omega_\BG + 3 \vson_\QUINT).
\end{equation}
Let us now concentrate on the homogeneous part of
Eq.~(\ref{KGlwave}). Using the previous equation, it can be expressed
as
\begin{eqnarray}
\label{KGlwhom}
 &   & \delta Q'' + \frac{4}{1 + 3 \omega_\BG} \frac{1}{\eta} \delta Q '
\nonumber \\
 & + & \frac{3}{(1 + 3 \omega_\BG)^2} \frac{1}{\eta^2} (1-\vson_\QUINT ) 
       (6 + 3 \omega_\BG + 3\vson_\QUINT) \delta Q = 0.
\end{eqnarray}
This linear equation can easily be solved: its solutions are just
power law of the conformal time. However, in order to show explicitly
the complete analogy with the background attractor, we choose to
analyse it in a rather roundabout way. Let us proceed exactly as for
the unperturbed Klein-Gordon equation [see the discussion around
Eq.~(\ref{KGmatrix})]. We define the time $\tau$ by $\eta \equiv
e^\tau$ and introduce the quantity $\delta u$ and $\delta p$ defined
by $\delta u \equiv \delta Q$ and $\delta p \equiv \ddd(\delta Q) /
\ddd \tau $. Then, Eq.~(\ref{KGlwhom}) can re-expressed as
\begin{eqnarray}
\label{attracpert}
& &   \frac{\ddd}{\ddd \tau} \pmatrix{\delta p \cr 
                                      \delta u}
\nonumber \\
& & = \pmatrix{\frac{3 (\omega_\BG - 1)}{1 + 3\omega_\BG} & 
               \frac{9 (\vson_\QUINT - 1)}{(1 + 3\omega_\BG)^2}
               (2 + \omega_\BG + \vson_\QUINT ) \cr 
               1 & 0}
      \pmatrix{\delta p \cr
               \delta u}. 
\end{eqnarray}
The form of this equation clearly shows the complete analogy with
Eq.~(\ref{KGmatrix}). The eigenvalues of the system are found by
solving the equation $\mbox{det}(M - \lambda I) = 0$, where $M$ is the
matrix defined above and $I$ the identity matrix. Straigthforwards
calculations show that the solutions are given by
\begin{equation}
\label{sollambda}
\lambda_\pm = \frac{3}{2} \frac{\omega_\BG - 1}{1 + 3\omega_\BG}
              \biggl[1 \pm \sqrt{1 + 4 \frac{\vson_\QUINT - 1}
                                           {(\omega_\BG - 1)^2}
                                     (2 + \omega_\BG + \vson_\QUINT)}
              \biggr].
\end{equation}
Of course, this is just a simple rephrasing of the fact that the
solution of Eq.~(\ref{KGlwhom}) is $\delta Q \propto
A_+\eta^{\lambda_+} + A_-\eta^{\lambda_-}$. The presence of an
attractor is linked to the negative sign of the real part of
$\lambda_\pm$. It is easy to see that the real part is always negative
in all four regimes, in particular this is true for any value of
$\alpha$. This is displayed in Fig.~\ref{toto} in the plane
$(\omega_\BG, \vson_\QUINT)$. The green and purple regions are the
regions where these real parts are negative. The green region is the
region where the argument of the square root is negative, \IE{} where
the square root is an imaginary number. The exact ``trajectories'' of
the system for the usual tracking potential (short line) and for the
SUGRA tracking potential (long line) are also shown for the case
$\alpha = 11$. They have been obtained by full numerical
integration. The remarkable property is that these trajectories are
always in the stable region. This means that, in each region, the
system tends to an attractor which is given by the inhomogeneous part
of the perturbed Klein-Gordon equation. The system starts at
$\omega_\BG = 1 / 3$ and goes from $\vson_\QUINT = 1$ to $\vson_\QUINT
= - 7 /3$. Then, the system approaches the transition to the matter
dominated era and leaves the vertical line. Finally, it stops when the
redshift vanishes at $\omega_\BG \simeq -0.29$ for the tracking
potential and at $\omega_\BG \simeq -0.82$ for the SUGRA tracking
potential. The two lines separate when the exponential factor becomes
important in the SUGRA tracking potential.

The conclusion is that the final value of the quintessence
perturbations is insensitive to the initial conditions, a property
completely similar to what has been shown in Ref.~\cite{SWZ} for the
background. Strictly speaking, this property has been demonstrated for
long wavelength modes only. However, we have checked by numerical
calculations that this is also true for shorter wavelength
modes. Having proven that the final result does not depend on the
initial conditions of the quintessence perturbations, we can now
proceed further and embark in a rather detailed study of the CMB 
anisotropies predictions in the presence of quintessence.

\section{Predictions for the power spectrum and the multipole moments}
\label{sec_pred_cl}

The presence of cosmological perturbations induces directional
variations in the CMB photon redshift. This is the so-called
Sachs-Wolfe effect~\cite{SWolfe}. Since these variations are the same
regardless of the wavelength of the photons, they translate into
variations in the temperature of the black body on the celestial
sphere. Their amplitude has been measured by the COBE satellite and is
of the order of magnitude $\delta T / T_0 \simeq
10^{-5}$~\cite{COBE}. The detailed angular structure of the CMB
anisotropies is usually characterized by the two-point correlation
function which can be expanded according to
\begin{equation}
\label{2pts}
\biggl \langle \frac{\delta T}{T}({\bf e}_1)
               \frac{\delta T}{T}({\bf e}_2)
\biggr \rangle
 = \frac{1}{4 \pi} \sum_l (2 l + 1) C_l P_l(\cos \gamma),
\end{equation} 
where $\gamma$ is the angle between the directions ${\bf e}_1$ and
${\bf e}_2$ and $P_l$ is a Legendre polynomial. The coefficients
$C_l$ are the multipole moments. In what follows, we will be mainly
interested in the so-called band power $\delta T_l$ defined by the
following expression
\begin{equation}
\label{defdtl}
\delta T_l \equiv T_0 \sqrt{l (l + 1) \frac{C_l}{2 \pi}},
\end{equation}
where $T_0\simeq 2.7 \UUNIT{K}{}$. The band power has now been
measured on a wide range of angular scales from $10'$ to $90^\circ$
corresponding roughly to $l \in [2, 700]$. Almost $80$ data points
have been measured. Recently new data obtained by the balloon-borne
experiments BOOMERanG~\cite{B98a,B98b} and MAXIMA-1~\cite{MAX1a,MAX1b}
have been published. They clearly show a detection of the first
Doppler peak at the expected angular scale $\simeq 1^\circ$
corresponding to the size of the Hubble radius at recombination.
\par
On the theoretical side, the multipoles moments depend on the initial
spectra for scalar and tensor modes and on how the perturbations
evolve from the initial time (after inflation) until now. This
evolution is determined by the values of the cosmological parameters,
\IE{} by the value of the Hubble constant ($h$), of the total amount
of matter present in our Universe ($\Omega_0$), of the cosmological
constant ($\Omega_\Lambda$), of the baryons density parameter 
($\Omega_\BAR$) and of the cold dark matter density parameter 
($\Omega_\CDM$). Constraints already exist on some of these
parameters. In particular, as already mentioned above, $\Omega_\Lambda
\simeq 0.7$ according to the SNIa measurements and $h^2 \Omega_\BAR
\simeq 0.019\pm 0.002$ according to BBN~\cite{Ol,NSB}. We also assume
$\Omega_0 = 1$ in agreement with the inflation paradigm which has been
confirmed by the recent CMB anisotropy measurements. For the
initial spectra, it is traditional to assume that they are of the
power-law form
\begin{equation}
\label{inispec}
k^3 P_\Phi    (k) = A_\SCAL k^{n_\SCAL - 1}, \quad 
k^3 P_{\rm h} (k) = A_\TENS k^{n_\TENS},
\end{equation}
where the scalar and tensor spectral indices $n_\SCAL$ and $n_\TENS$
are related by $n_\SCAL - 1 = n_\TENS$. This last equation is also
valid for zeroth order slow-roll inflation. It should be noticed that,
{\it a priori}, this choice is not the most relevant one since
slow-roll inflation is certainly more physically motivated. For
spectral indices close to $n_\SCAL = 1$, we expect a small
difference. This is no longer true for larger tilts. Inflation
predicts the presence of gravitational perturbations and the tensor to
scalar amplitude ratio is given by
\begin{equation}
\label{TSratio}
\frac{A_\TENS}{A_\SCAL} \simeq - \frac{200}{9} n_\TENS.
\end{equation}
This equation is valid for power-law inflation with $n_\TENS$ not too
large\footnote{For power-law inflation, the exact expression is given
by $A_\TENS / A_\SCAL = - (200 / 9) n_\TENS / (1 - n_\TENS / 2)$.} or
for zeroth order slow-roll inflation. A last remark is in order at
this point. All the plots displayed in this article are COBE
normalized in the following way: the position of the Sachs-Wolfe
plateau is tuned such that it best fits the COBE data points. In
practice, this almost amounts to normalize the spectrum to $C_{10}$.
\par
In this section, we first study the general properties of the
multipoles moments in the quintessence cold dark matter model (QCDM)
and point out the main differences with the standard cold dark matter
(sCDM) and the cosmic concordance model ($\Lambda$CDM). We also
display the corresponding baryonic matter power spectra, given by
\begin{equation}
\label{matterps}
\vert \delta (k) \vert ^2 
\equiv \biggl \vert \frac{\delta \rho_\BAR}{\rho_\BAR} \biggr \vert^2, 
\end{equation}
which is the square of the Fourier transform of the baryonic density
contrast. Then, we compare the predictions of the QCDM model for the
Ratra-Peebles and SUGRA tracking potentials with the COBE~\cite{COBE},
BOOMERanG~\cite{B98a,B98b}, MAXIMA-1~\cite{MAX1a,MAX1b} and
Saskatoon~\cite{Sask} data. We do not attempt to perform a 
detailed statistical analysis but we rather indicate roughly how
the different models can fit the observational data.
\par
\begin{figure}
\centerline{\psfig{file=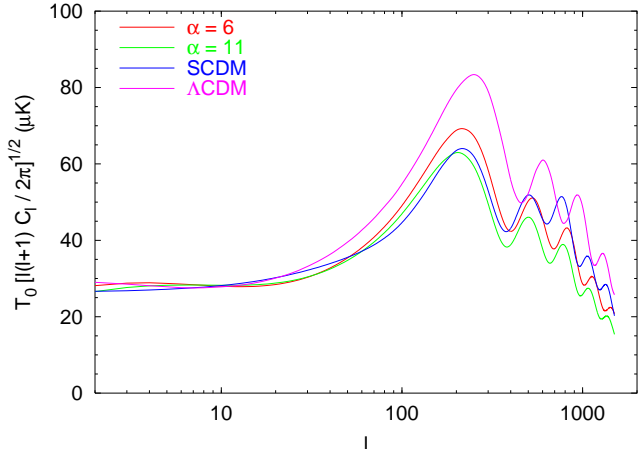,width=3.5in,angle=-90}}
\caption{Multipole moments for the Ratra-Peebles potential for
two values of $\alpha$, $\alpha = 6$ (red curve) and $\alpha = 11$
(green curve) and with cosmological parameters equal to $h = 0.5$,
$\Omega_\BAR = 0.05$, $\Omega_\QUINT = 0.7$, $\Omega_\CDM = 1 -
\Omega_\BAR - \Omega_\QUINT$, $n_\SCAL=0.99$, $A_\TENS=0$. The curves
are compared with those obtained in the SCDM model (blue curve)
and in the $\Lambda$CDM model (pink curve).}
\label{Trp}
\end{figure}
\begin{figure}
\centerline{\psfig{file=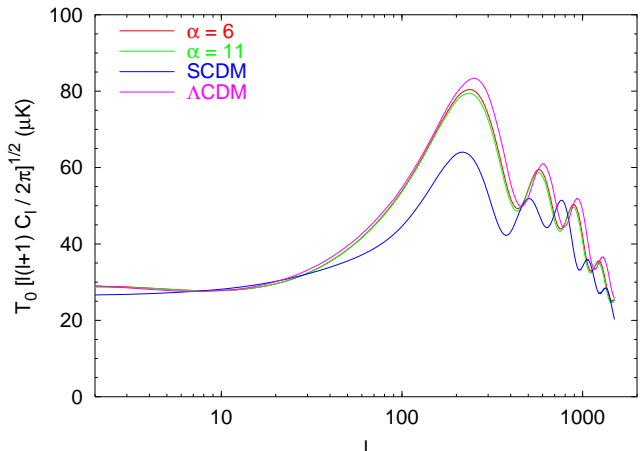,width=3.5in,angle=-90}}
\caption{Multipole moments for the SUGRA potential for two values
of $\alpha$, $\alpha = 6$ (red curve) and $\alpha = 11$ (green curve)
and with cosmological parameters equal to $h = 0.5$, $\Omega_\BAR =
0.05$, $\Omega_\QUINT = 0.7$, $\Omega_\CDM = 1 - \Omega_\BAR -
\Omega_\QUINT$, $n_\SCAL = 0.99$, $A_\TENS = 0$. The curves are
compared with those obtained in the SCDM model (blue curve) and in
the $\Lambda$CDM model (pink curve).}
\label{Tsugra}
\end{figure}

We now turn to simple considerations about the shape of the CMB
spectrum. The corresponding band power for the Ratra-Peebles and SUGRA
potentials are displayed in Figs.~\ref{Trp} and~\ref{Tsugra} for $h =
0.5$, $\Omega_\BAR = 0.05$, $\Omega_\QUINT = 0.7$, $\Omega_\CDM
= 1 - \Omega_\BAR - \Omega_\QUINT$, $n_\SCAL = 0.99$ and the tensor
contribution neglected. The former set of cosmological parameters has
been chosen just for the sake of illustration and discussion.
For simplicity, we start with a comparison of the quintessence
multipole moments with those obtained in the $\Lambda$CDM model with
similar cosmological parameters. Firstly, since $\Omega_\MAT \equiv
\Omega_\CDM + \Omega_\BAR$ is the same in the two models, the redshift
of equivalence between matter and radiation $z_{\rm eq} \equiv
\Omega_\MAT / \Omega_\RAD$, where $\Omega_\RAD \equiv \Omega_\gamma +
\Omega_\nu$, is also the same in both cases. Therefore, the first peak
is boosted in the same way by the early integrated Sachs-Wolfe effect
(due to the time variation of the two Bardeen potentials during
recombination, see~\cite{hu95}) and, {\it a priori}, one expects the
same first peak height. Secondly, the dark energy component
(cosmological constant or quintessence) has a negligible contribution
before recombination and, as a consequence, the evolution of the
perturbations before the last scattering surface is the same in the
two models (see the previous section). Thus, one expects again
identical acoustic peak patterns. However, despite the previous
considerations, the position of the peaks differs because the angular
distance-redshift relation is modified at small redshift since the
equation of state of the cosmological constant and of quintessence is
not the same. The closest to $-1$ the equation of state parameter is,
the largest the shift of the peaks to small angular scales is. As a
consequence, the peaks in the $\Lambda$CDM model are more shifted to
the right than in the QCDM model. Another feature is that the height
of the first peak is not the same in the two types of
scenarios. Indeed, at small redshift, the gravitational potential does
not behave exactly in the same way in the two models especially
because there are scalar field perturbations in the QCDM
scenario. This results in a different contribution of the late
integrated Sach-Wolfe effect~\cite{hu95} which affects the overall
normalization of the spectrum. As a consequence, the height of the
first peak is lower in the model which produces a strong late
integrated Sachs-Wolfe effect, \IE{} in the QCDM model.
\par
The exact shape of the quintessence potential also matters and
different potentials lead to different CMB anisotropies. The SUGRA
potential and the cosmological constant lead to very similar CMB
anisotropy spectra, whereas the difference is stronger in the case of
the Ratra-Peebles potential. This is mainly due to the fact that the
equation of state parameter is generically closer to $-1$ in the first
case than in the second one. Another difference is that the
Ratra-Peebles potential produces a larger late integrated Sachs-Wolfe
contribution than the SUGRA potential. This results in a different
normalization for both models (note that the normalisation depends on
$\alpha$) which has for consequence different height of the first
Doppler peak. Of course, this difference is also visible in the power
spectrum at large scales. Maybe the most interesting property is the
following one. The cosmic equation of state (almost) does not depend
on $\alpha$ in the case of the SUGRA potential. Then, in the same
manner, the CMB anisotropies do not depend on $\alpha$ contrary to the
case of the Ratra-Peebles potential. This means that the multipole
moments displayed in Fig.~\ref{Tsugra} are a generic predictions of
the SUGRA QCDM model.
\par
\begin{figure}
\centerline{\psfig{file=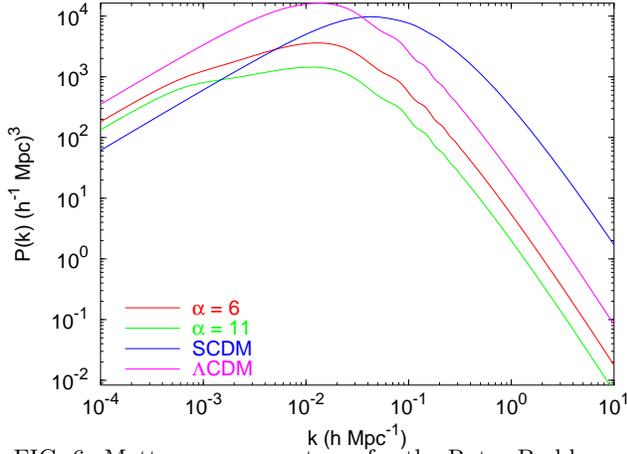,width=3.5in,angle=-90}}
\caption{Matter power spectrum for the Ratra-Peebles potential for
two values of $\alpha$, $\alpha = 6$ (red curve) and $\alpha = 11$
(green curve) and with cosmological parameters equal to $h = 0.5$,
$\Omega_\BAR = 0.05$, $\Omega_\QUINT = 0.7$, $\Omega_\CDM = 1 -
\Omega_\BAR - \Omega_\QUINT$, $n_\SCAL = 0.99$, $A_\TENS = 0$. The
curves are compared with those obtained in the SCDM model (blue
curve) and in the $\Lambda$CDM model (pink curve).}
\label{Prp}
\end{figure}
\begin{figure}
\centerline{\psfig{file=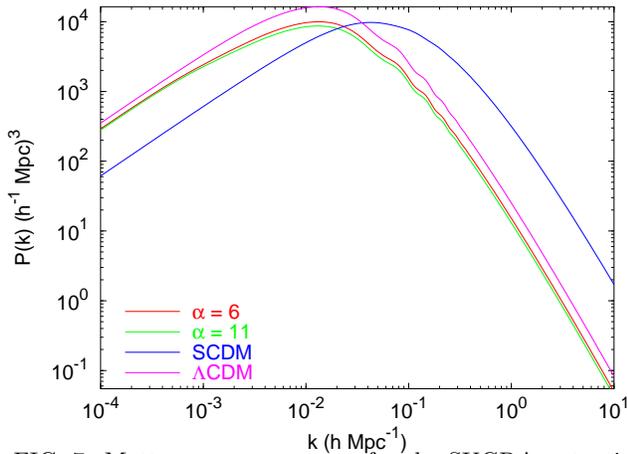,width=3.5in,angle=-90}}
\caption{Matter power spectrum for the SUGRA potential for 
two values of $\alpha$, $\alpha = 6$ (red curve) and $\alpha = 11$
(green curve) and with cosmological parameters equal to $h = 0.5$,
$\Omega_\BAR = 0.05$, $\Omega_\QUINT = 0.7$, $\Omega_\CDM = 1 -
\Omega_\BAR - \Omega_\QUINT$, $n_\SCAL = 0.99$, $A_\TENS = 0$. The
curves are compared with those obtained in the SCDM model (blue
curve) and in the $\Lambda$CDM model (pink curve).}
\label{Psugra}
\end{figure}

For the sake of completness, let us now describe the corresponding
matter power spectra. They are displayed in Fig.~\ref{Prp}
and~\ref{Psugra}.
The matter power spectrum also depends on the nature of the dark
energy component (cosmological constant or quintessence) but the
difference between the cosmological constant scenario and a
quintessence scenario is less important. The matter power spectrum
shows a peak the location of which is given by the Hubble radius at
equivalence. In the $\Lambda$CDM and QCDM scenarios, the peak is at
the same location contrary to the sCDM case for which the peak is
located at smaller scales. Also, in models with low matter content,
the ratio $\Omega_\BAR / \Omega_\CDM$ is higher which results in the
presence of smooth oscillations at small scales. As for the CMB
anisotropy spectrum, the small scales are similar in the $\Lambda$CDM
and QCDM scenarios and important differences only occur on larger
scales which are more affected by the change in the cosmic equation of
state.
\par
\begin{figure}
\centerline{\psfig{file=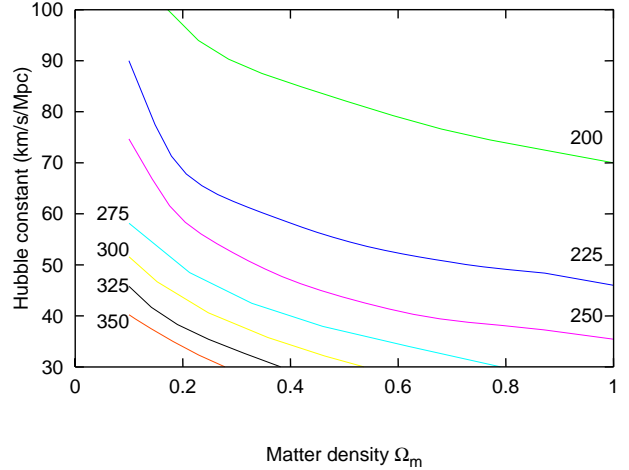,bbllx=135pt,bblly=150pt,bburx=500pt,bbury=630pt,width=3.5in,angle=-90}}
\caption{Contour plots of the first Doppler peak location in the 
$(\Omega_\MAT, h)$ plane for the cosmological constant case. The
other cosmological parameters are $h^2 \Omega_\BAR = 0.019$, 
$\Omega_\Lambda = 0.7$, $A_\TENS = 0$ and $n_\SCAL = 0.99$.}
\label{l1L}
\end{figure}
\begin{figure}
\centerline{\psfig{file=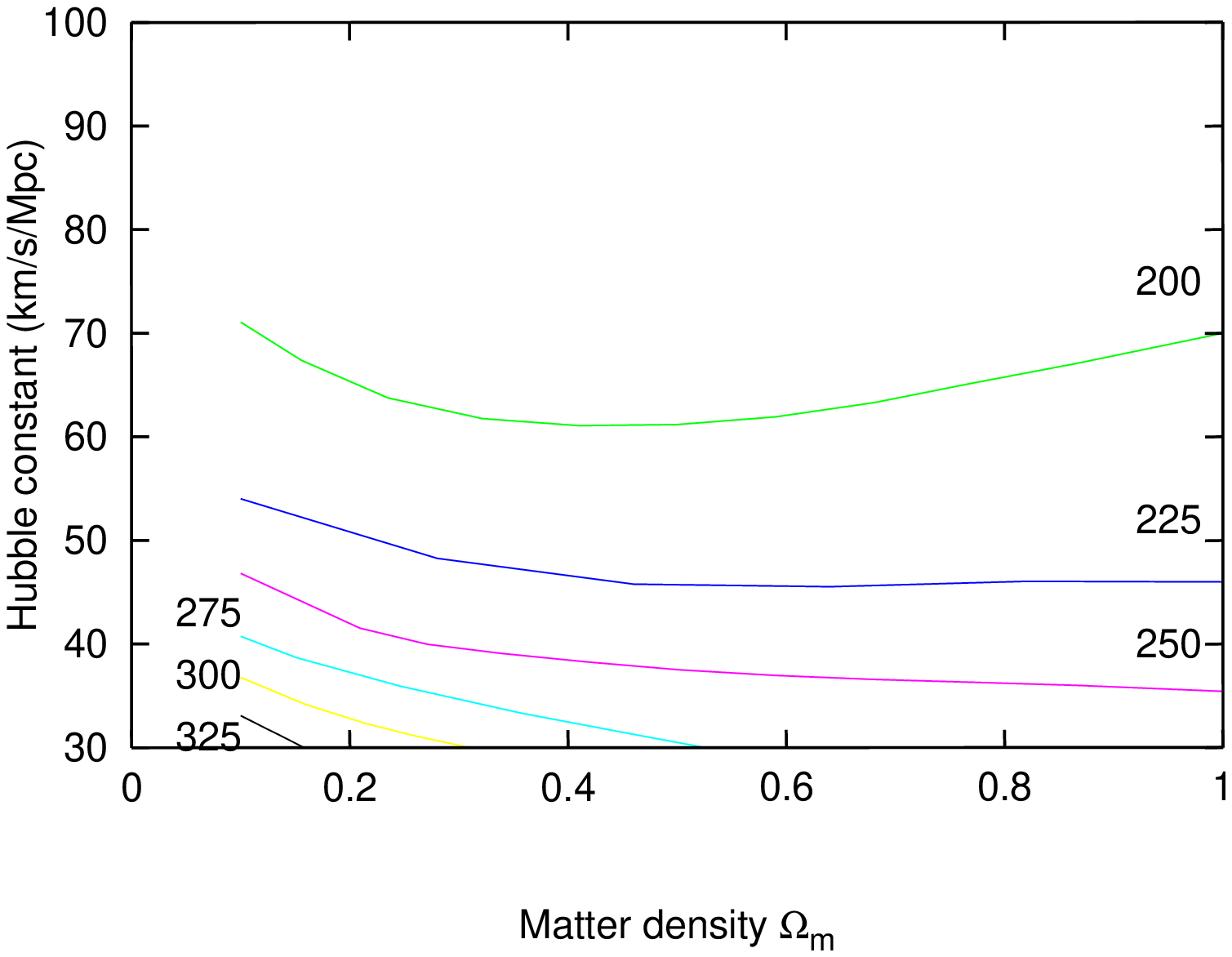,bbllx=135pt,bblly=150pt,bburx=500pt,bbury=630pt,width=3.5in,angle=-90}}
\caption{Contour plots of the first Doppler peak location in the 
$(\Omega_\MAT, h)$ plane for the Ratra-Peebles QCDM case. The
other cosmological parameters are $h^2 \Omega_\BAR = 0.019$, 
$\Omega_\QUINT = 0.7$, $A_\TENS = 0$ and $n_\SCAL = 0.99$.}
\label{l1rp}
\end{figure}
\begin{figure}
\centerline{\psfig{file=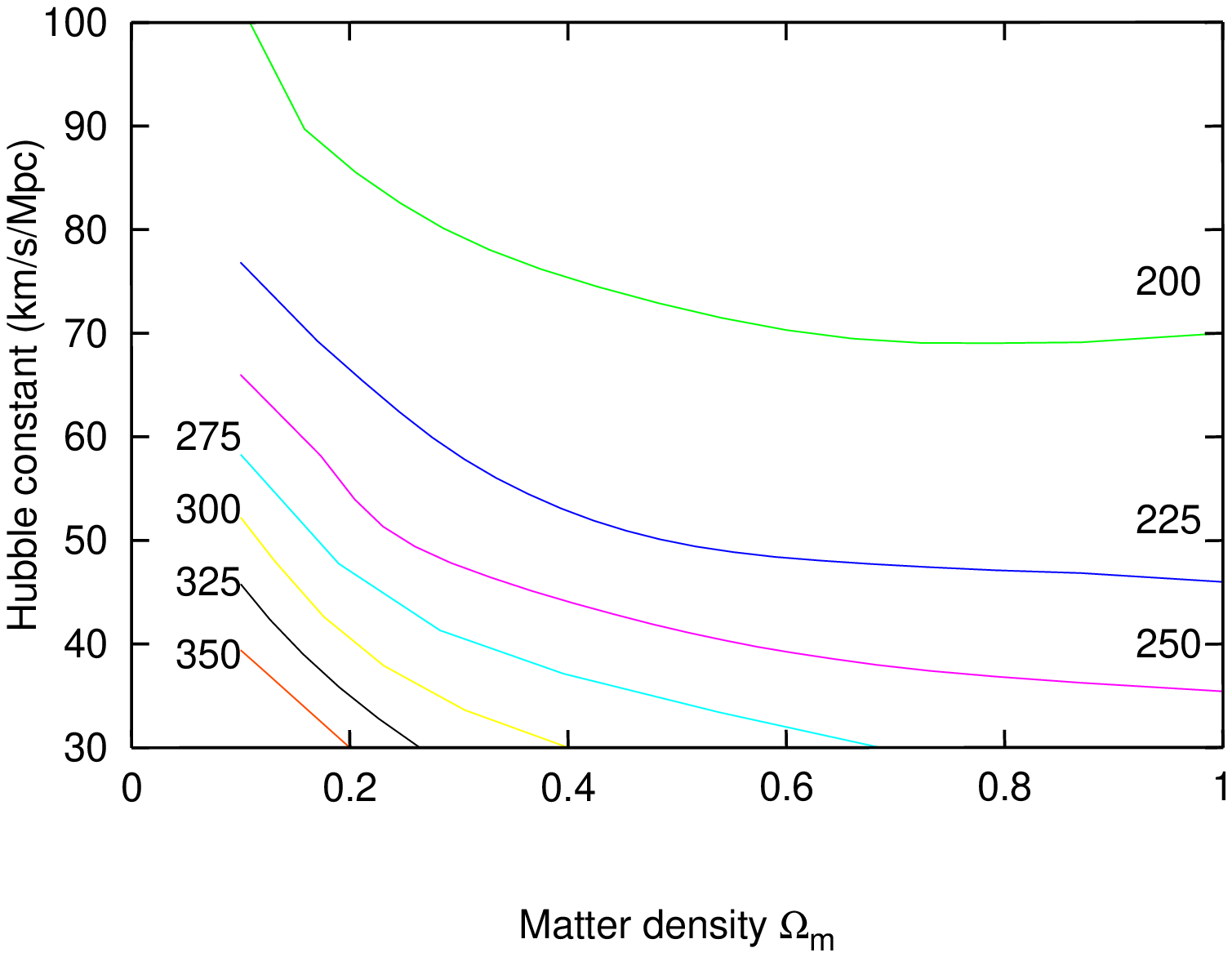,bbllx=135pt,bblly=150pt,bburx=500pt,bbury=630pt,width=3.5in,angle=-90}}
\caption{Contour plots of the first Doppler peak location in the 
$(\Omega_\MAT, h)$ plane for the SUGRA QCDM case. The other
cosmological parameters are $h^2 \Omega_\BAR = 0.019$, 
$\Omega_\QUINT = 0.7$, $A_\TENS = 0$ and $n_\SCAL = 0.99$.}
\label{l1sugra}
\end{figure}

Let us now study in more details and for more realistic values of the
cosmological parameters, the position and the height of the first
Doppler peak. We start with the location of the first peak (denoted in
what follows by $l_1$) and we study it in the plane $(\Omega_\MAT,
h)$ with the following values of the other cosmological parameters:
$h^2 \Omega_\BAR = 0.019$ (the value predicted by standard BBN),
$\Omega_{\Lambda, \QUINT} = 0.7$, $A_\TENS = 0$ and $n_\SCAL =
0.99$. The case of the cosmological constant is displayed in
Fig.~\ref{l1L}, the case of the Ratra-Peebles QCDM model in
Fig.~\ref{l1rp} and the case of the SUGRA QCDM model in
Fig.~\ref{l1sugra}.
These plots confirm the qualitative predictions made above and in
particular the fact that, in general, $l_1^\Lambda > l_1^\SUGRA >
l_1^\RP$. If one assumes that $\Omega_\MAT \simeq 0.3$ (since we have
assumed $\Omega_\Lambda \simeq 0.7$) and $h \simeq 0.62$, this last
value being consistent with the Hubble Space Telescope (HST) and SNIa
measurements, then we obtain $l_1^\Lambda \simeq 225$, $l_1^\SUGRA
\simeq 220$ and $l_1^\RP\simeq 200$. It is interesting to compare
these values with the recent measurements of the first peak performed
by BOOMERanG and MAXIMA-1. The BOOMERanG data indicate that
$l_1=197\pm 6$~\cite{B98a,B98b} which is compatible with the
Ratra-Peebles potential and a spatially flat Universe. On the other
hand, the MAXIMA-1 data are consistent with a first peak located at
$l_1\simeq 220$~\cite{MAX1a,MAX1b} which is, this time, in agreement
with a cosmological constant or the SUGRA QCDM model.
\par
\begin{figure}
\centerline{\psfig{file=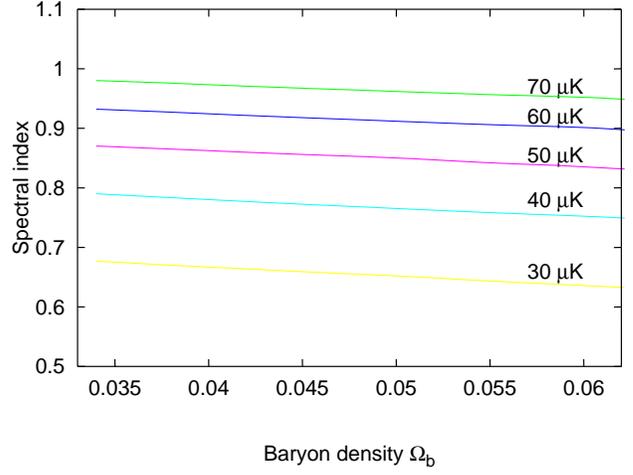,bbllx=60pt,bblly=180pt,bburx=500pt,bbury=750pt,width=3.5in,angle=-90}}
\caption{Contour plots of the height of the first peak in the 
$(\Omega_\BAR, n_\SCAL)$ plane with $h = 0.62$, $\Omega_\Lambda
= 0.7$ for the case of the $\Lambda$CDM model.}
\label{heiL}
\end{figure}
\begin{figure}
\centerline{\psfig{file=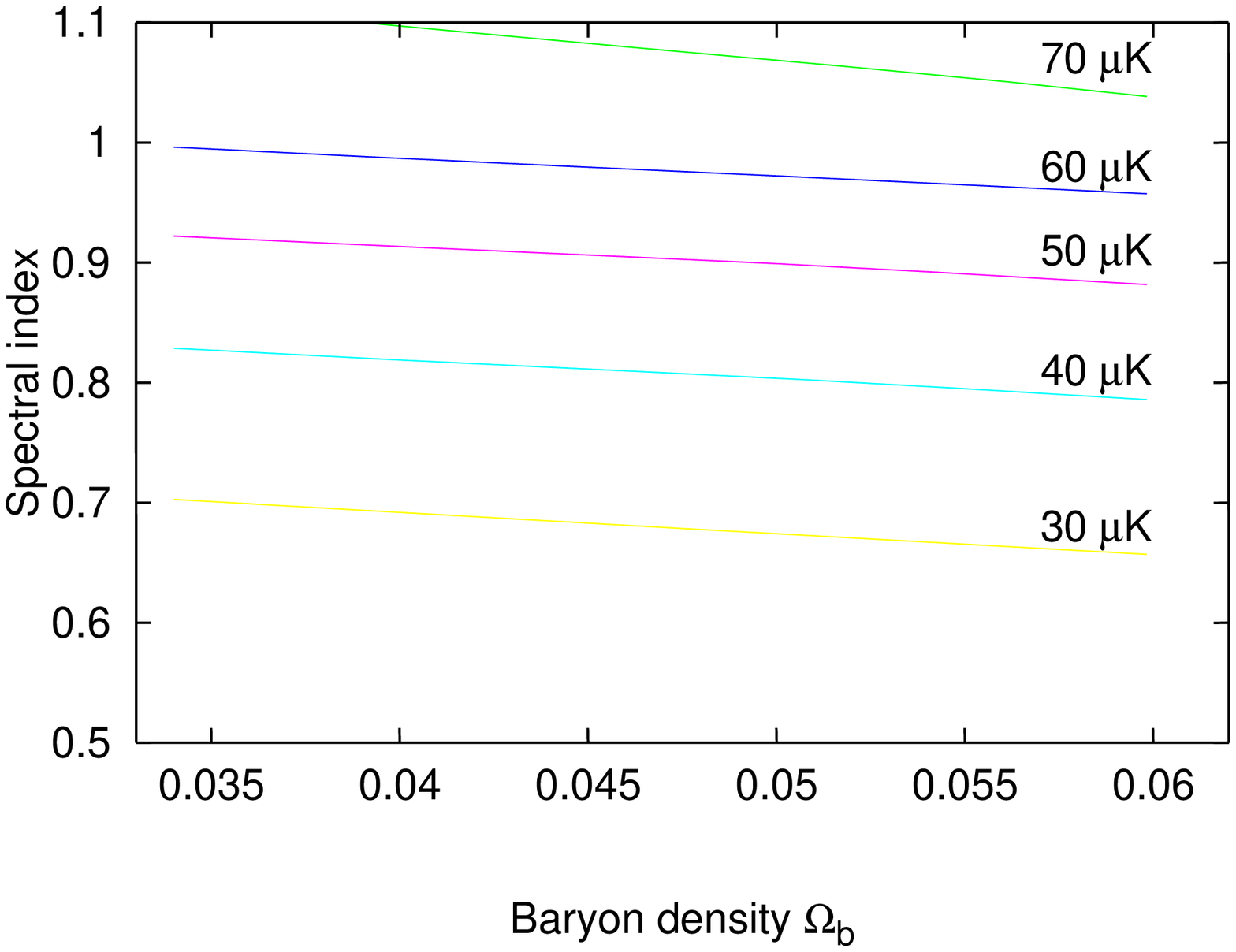,bbllx=60pt,bblly=180pt,bburx=500pt,bbury=750pt,width=3.5in,angle=-90}}
\caption{Contour plots of the height of the first peak in the 
$(\Omega_\BAR, n_\SCAL)$ plane with $h=0.62$, $\Omega_\QUINT =
0.7$ for the case of the Ratra-Peebles QCDM model.}
\label{heirp}
\end{figure}
\begin{figure}
\centerline{\psfig{file=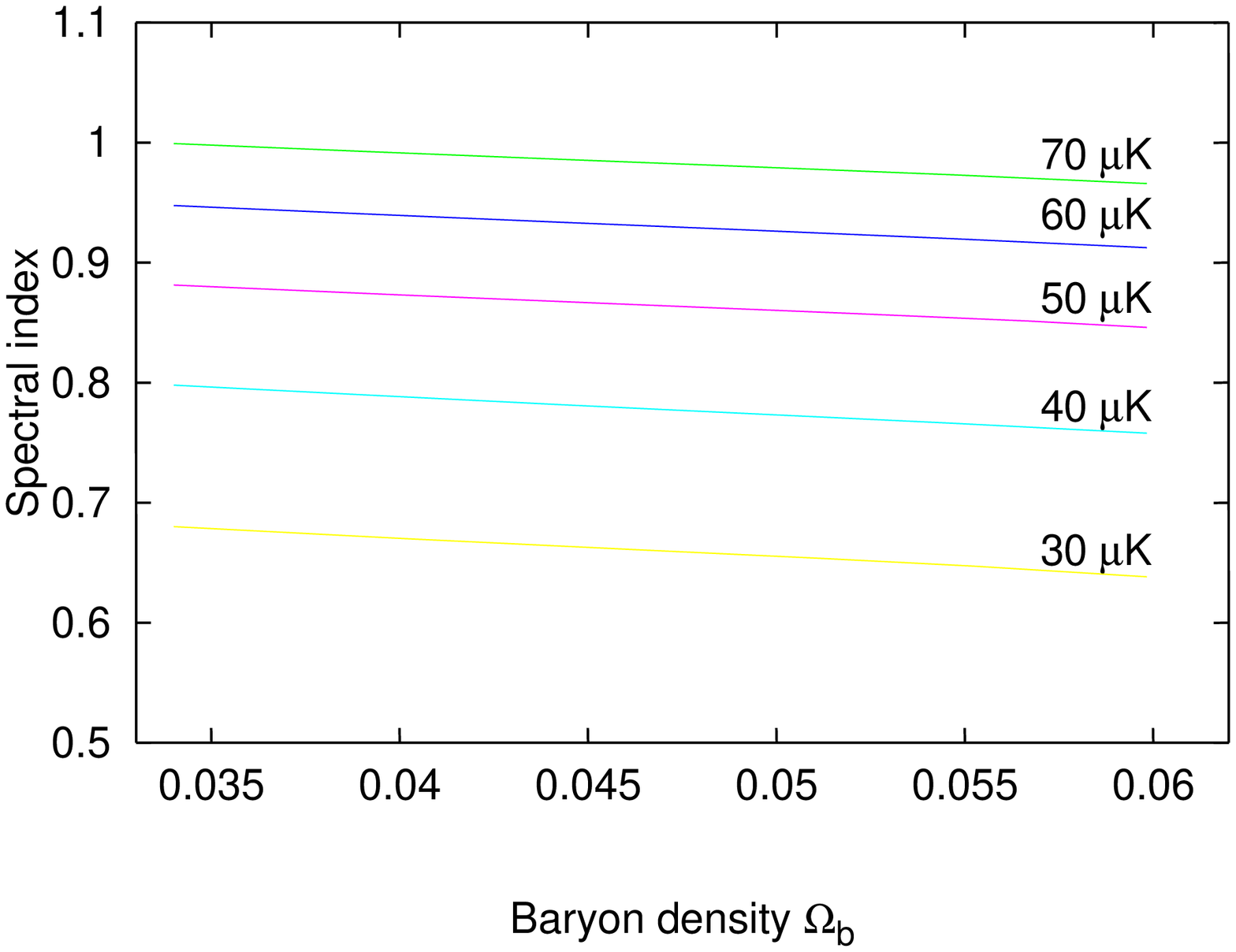,bbllx=60pt,bblly=180pt,bburx=500pt,bbury=750pt,width=3.5in,angle=-90}}
\caption{Contour plots of the height of the first peak in the 
$(\Omega_\BAR, n_\SCAL)$ plane with $h = 0.62$, $\Omega_\QUINT =
0.7$ for the case of the SUGRA QCDM model.}
\label{heisugra}
\end{figure}

Let us now study the height of the first Doppler peak. We study its
variation in the plane $(\Omega_\BAR, n_\SCAL)$ for the following
values of the cosmological parameters: $h = 0.62$, $\Omega_{\Lambda,
\QUINT} = 0.7$. The case of the $\Lambda$CDM model is displayed in
Fig.~\ref{heiL} whereas the cases of the Ratra-Peebles QCDM and SUGRA
QCDM are presented in Figs.~\ref{heirp} and~\ref{heisugra},
respectively.
We would like to emphasize that the importance of gravitational waves
is crucial in this case. Indeed, as already mentioned, the presence of
gravitational waves modifies the normalization and, as a consequence,
the height of the peaks. The BOOMERanG data indicate that $\delta
T_{200} \simeq 69 \pm 8 \UUNIT{\mu K}{}$~\cite{B98a,B98b} whereas the
MAXIMA-1 ones give $\delta T_{220} \simeq 78 \pm 6 \UUNIT{\mu
K}{}$~\cite{MAX1a,MAX1b}, this discrepancy being possibly explained by
problems in the calibration of these experiments. If we adopt the
value $\Omega_\BAR \simeq 0.0595$, compatible with BBN, we see that,
in the Ratra-Peebles QCDM model, a height of the first peak compatible
with the BOOMERanG and MAXIMA-1 data leads to a value of the scalar
spectral index such that $n_\SCAL > 1$. This is not compatible with
standard inflation and cannot be realized with one scalar field. We
interpret this as a new evidence that the Ratra-Peebles QCDM model (at
least with this value of $\alpha$) is excluded by the
observations. For the cases of $\Lambda$CDM and SUGRA QCDM, we learn
from the previous plots that the spectral index must be very close to
one.
\par
\begin{figure}
\centerline{\psfig{file=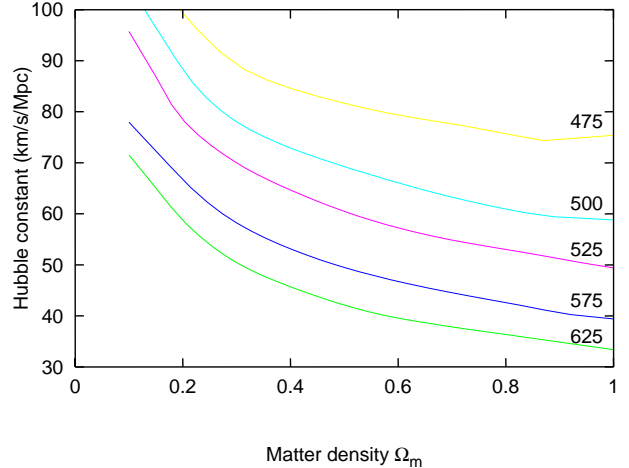,bbllx=135pt,bblly=150pt,bburx=500pt,bbury=630pt,width=3.5in,angle=-90}}
\caption{Contour plots of the location of the second peak in the 
$(\Omega_\MAT, h)$ plane with $h = 0.62$, $\Omega_\Lambda = 0.7$,
$h^2 \Omega_\BAR = 0.019$, $n_\SCAL = 0.99$ for the cosmological constant
case.}
\label{l2L}
\end{figure}
\begin{figure}
\centerline{\psfig{file=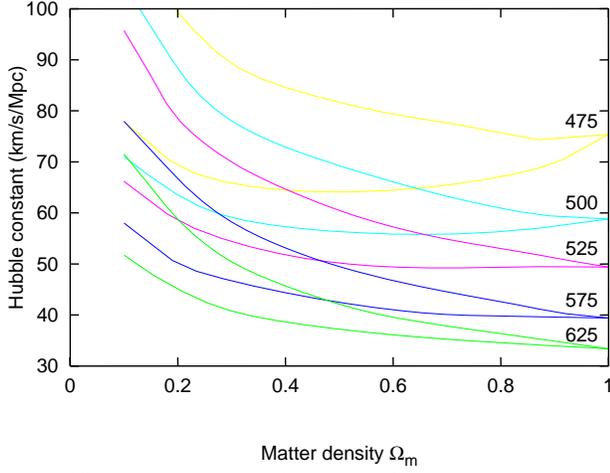,bbllx=135pt,bblly=150pt,bburx=500pt,bbury=630pt,width=3.5in,angle=-90}}
\caption{Contour plots of the location of the second peak in the 
$(\Omega_\MAT, h)$ plane with $h = 0.62$, $\Omega_{\Lambda,
\QUINT} = 0.7$, $h^2\Omega_\BAR = 0.019$, $n_\SCAL = 0.99$ for the
Ratra-Peebles QCDM model. The corresponding contour plots for the
cosmological constant (upper curves) are also displayed for
comparison.}
\label{l2rp}
\end{figure}
\begin{figure}
\centerline{\psfig{file=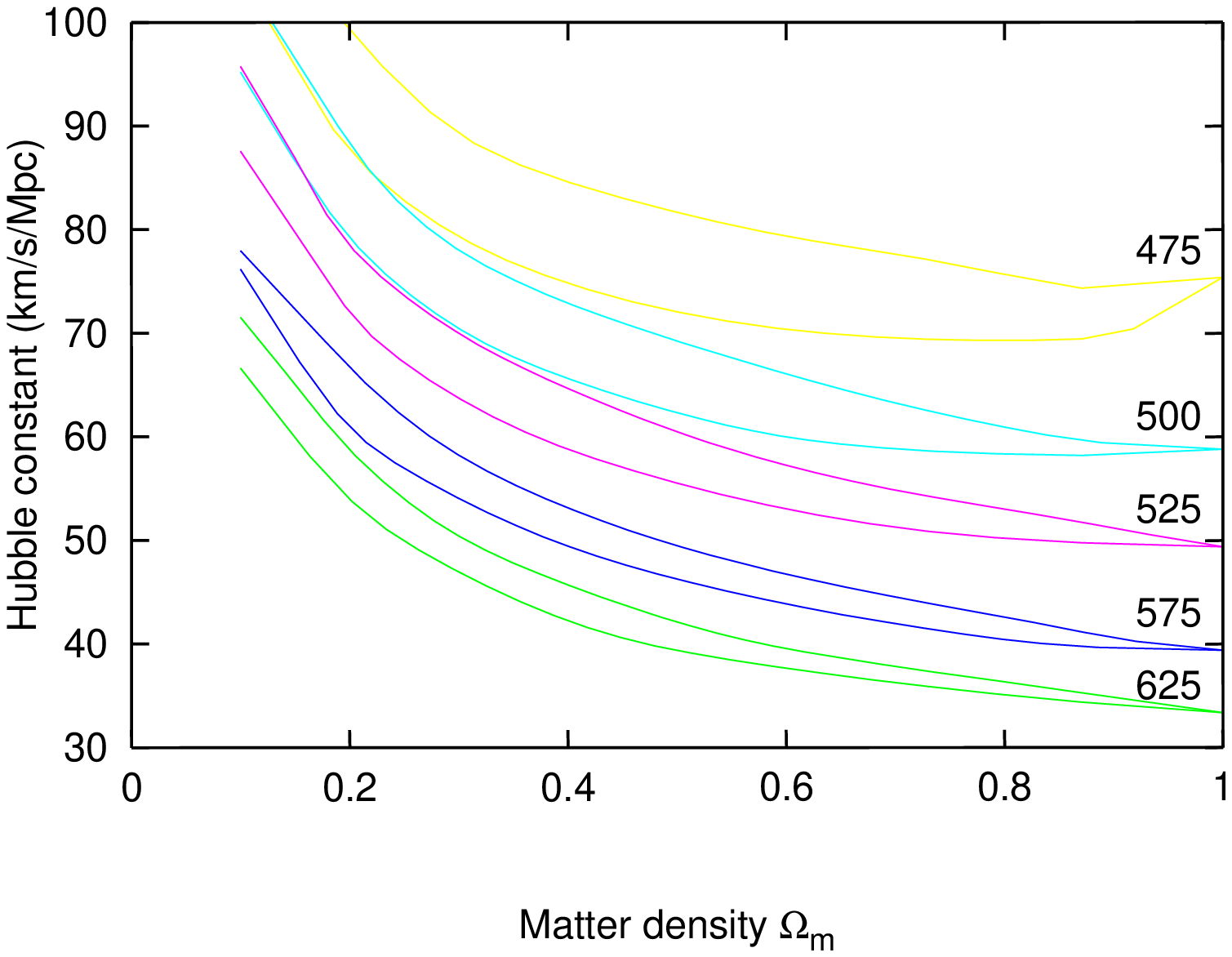,bbllx=135pt,bblly=150pt,bburx=500pt,bbury=630pt,width=3.5in,angle=-90}}
\caption{Contour plots of the location of the second peak in the 
$(\Omega_\MAT, h)$ plane with $h = 0.62$, $\Omega_{\Lambda,
\QUINT} = 0.7$, $h^2 \Omega_\BAR = 0.019$, $n_\SCAL = 0.99$ for the
SUGRA QCDM model. The corresponding contour plots for the cosmological
constant (upper curves) are also displayed for comparison.}
\label{l2sugra}
\end{figure}

We now turn to the study of the second Doppler peak. First of all, we
should say something about the observational situation. With regards
to the detection of a second peak, it is difficult to deduce something
from the BOOMERanG data. The error bars are still large and the data
are, for the moment, compatible with a second peak (with a height
maybe smaller than predicted by standard inflation) but also with no
peak at all, even if one can see a small rise of the signal at $l_2
\simeq 550$~\cite{B98a,B98b}. Only $5\%$ of the data of this
experimement have been analysed so far and one should wait for the
rest of the data analysis to be completed. On the other hand, the
MAXIMA-1 show ``a suggestion of a peak at $l_2 \simeq
525$''~\cite{MAX1a} the height of which would be $\delta T_{525}
\simeq 48 \UUNIT{\mu K}{}$. One could even argue that the
beginning of a third peak has been observed. In fact, considering all
the uncertainties in such measurements, we are of the opinion that a
reasonable attitute is simply to wait for more data.  On the
theoretical side, it was argued by Kamionkowski and
Buchalter~\cite{KB} that the location of the second peak can probe the
dark energy density. The main idea is to study the contour plots of
$l_2$ in the plane $(\Omega_\MAT, h)$. Then, a measurement of
$l_2$, knowing $h$ by other means , immediately determines the value
of $\Omega_\MAT$. It was claimed in Ref.~\cite{KB} that this strategy
does not depend on whether the dark energy is a cosmological constant
or a quintessence field. We show that this claim is not correct and
that the nature of the dark energy matters. The contour plots of $l_2$
in the case of a cosmological constant are displayed in Fig.~\ref{l2L}
for the cosmological parameters given by $h = 0.62$, $\Omega_\Lambda =
0.7$, $h^2 \Omega_\BAR = 0.019$, $n_\SCAL = 0.99$. These plots are in
agreement with the results found in Ref.~\cite{KB}.
The corresponding contour plots for the Ratra-Peebles and SUGRA CDM
models are presented in Figs.~\ref{l2rp} and~\ref{l2sugra}. In
addition, in order to show that there is indeed an important
difference, we also display the contour plots for a cosmological
constant which, for a given value of $l_2$, is always above the QCDM
curve. The fact that there is a difference does not totally invalidate
the idea of Ref. \cite{KB}. But it means that, in order to use it, we
should first identify the physical nature of the dark energy, for
example with a measurement of its equation of state parameter.
\begin{figure}
\centerline{\psfig{file=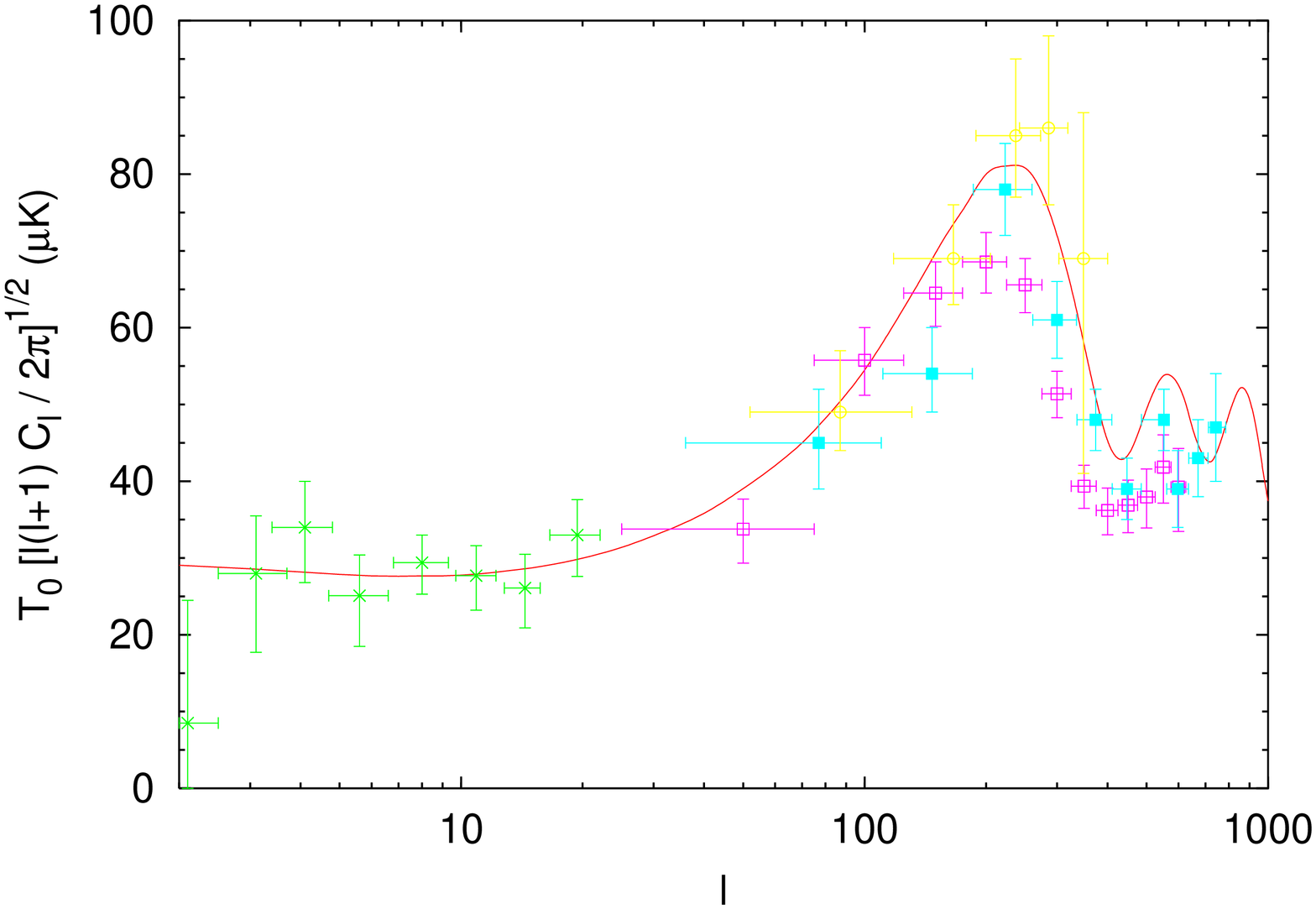,width=3.5in,angle=-90}}
\caption{Band power $\delta T_l$ for the $\Lambda$CDM model with
$h = 0.62$, $\Omega_\Lambda = 0.7$, $\Omega_\BAR = 0.595$ and $n_\SCAL
= 0.99$. The data point are those of the COBE, BOOMERanG, MAXIMA-1 and
Saskatoon experiments.}
\label{clL}
\end{figure}
\begin{figure}
\centerline{\psfig{file=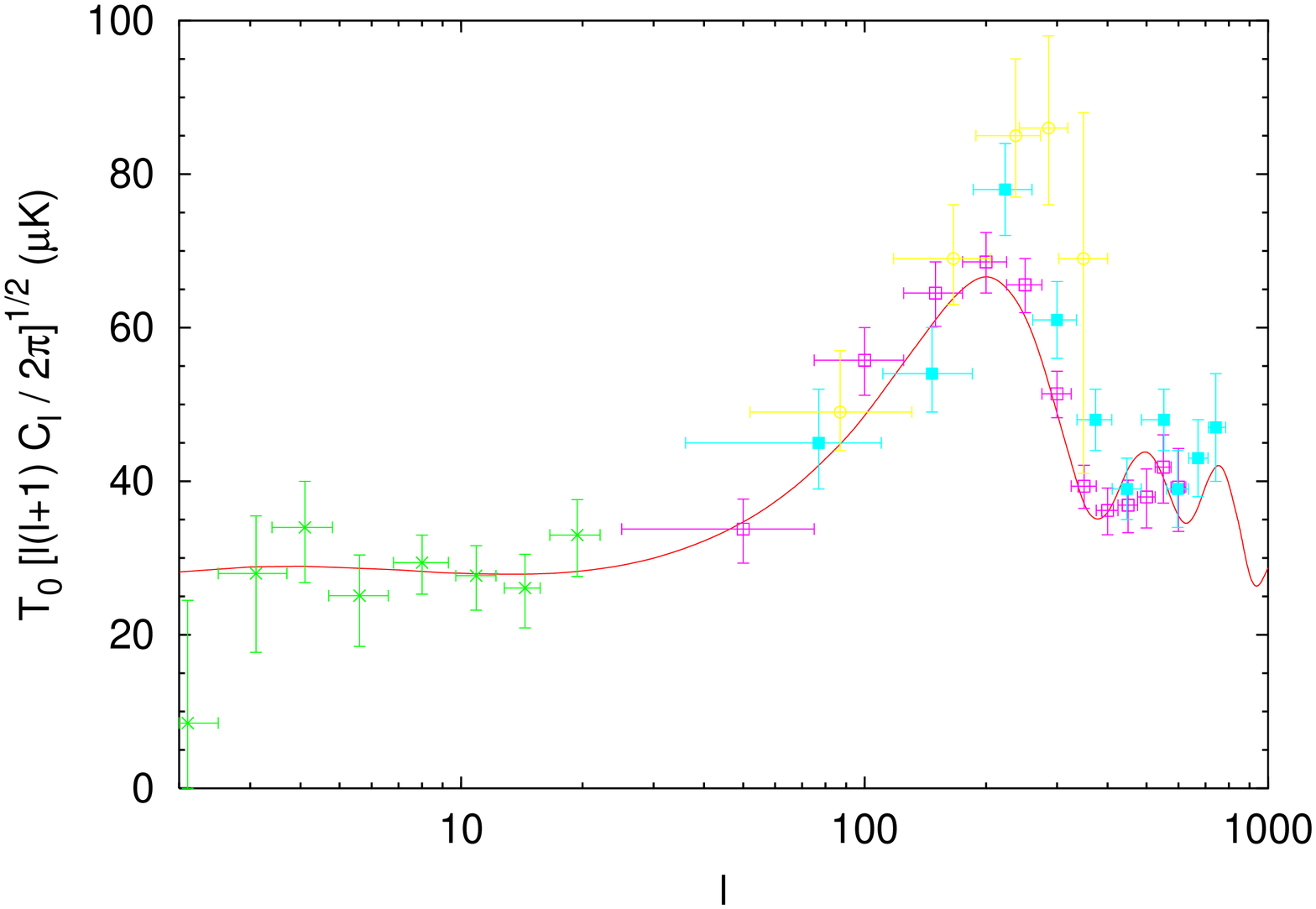,width=3.5in,angle=-90}}
\caption{Band power $\delta T_l$ for the Ratra-Peebles QCDM model
with $h = 0.62$, $\Omega_\QUINT = 0.7$, $\Omega_\BAR = 0.0595$ and
$n_\SCAL = 0.99$.  The data point are those of the COBE, BOOMERanG,
MAXIMA-1 and Saskatoon experiments.}
\label{clrp}
\end{figure}
\begin{figure}
\centerline{\psfig{file=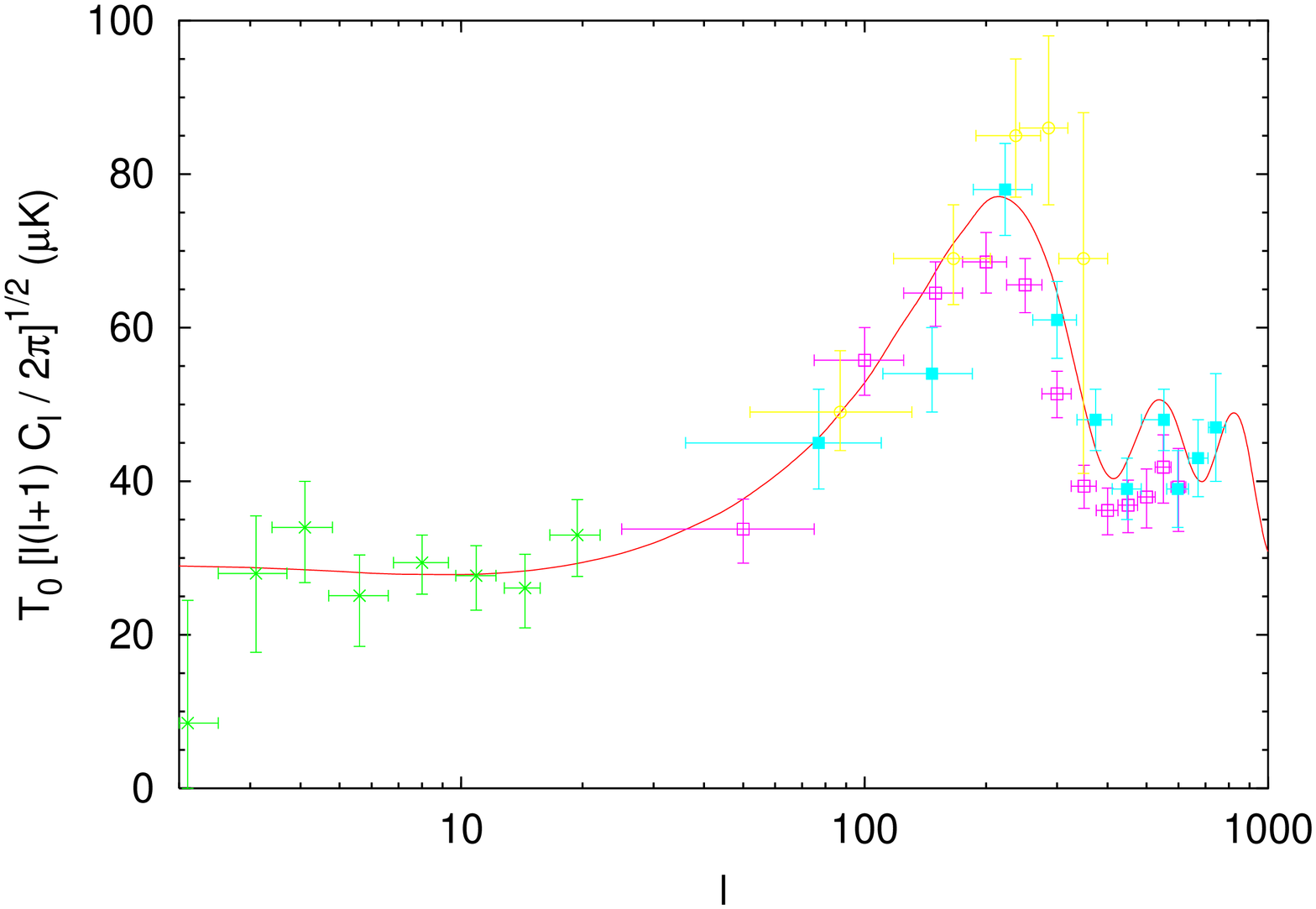,width=3.5in,angle=-90}}
\caption{Band power $\delta T_l$ for the SUGRA QCDM model with $h =
0.62$, $\Omega_\QUINT = 0.7$, $\Omega_\BAR = 0.0595$ and $n_\SCAL
= 0.99$.  The data point are those of the COBE, BOOMERanG, MAXIMA-1
and Saskatoon experiments.}
\label{clsugra}
\end{figure}
\begin{figure}
\centerline{\psfig{file=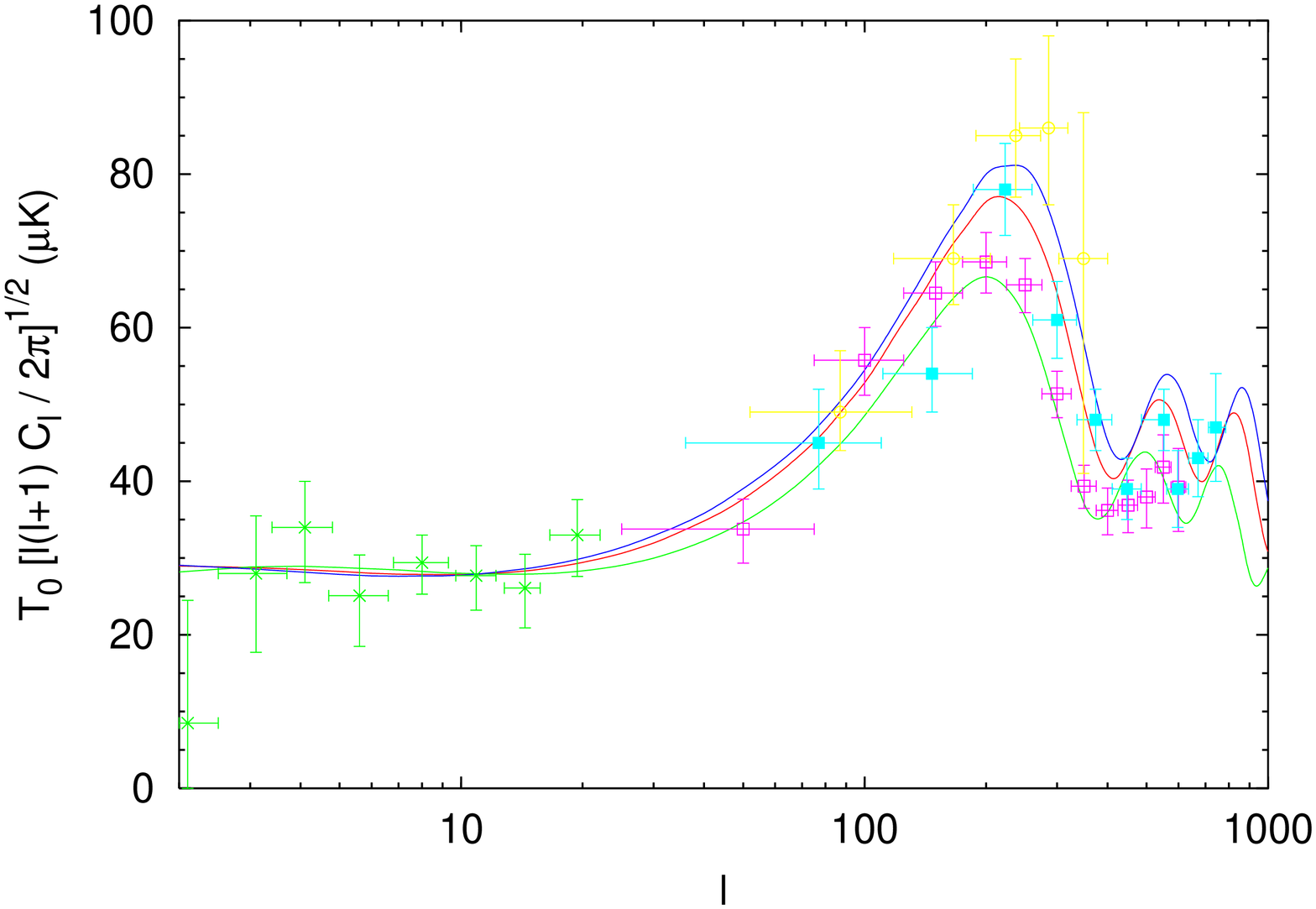,width=3.5in,angle=-90}}
\caption{Band power $\delta T_l$ for the $\Lambda$CDM model (blue
curve), Ratra-Peebles QCDM model (green curve) and the SUGRA QCDM
model (red curve) with $h=0.62$, $\Omega_{\Lambda, \QUINT} = 0.7$,
$\Omega_\BAR = 0.0595$ and $n_\SCAL = 0.99$.  The data point are those
of the COBE, BOOMERanG, MAXIMA-1 and Saskatoon experiments.}
\label{clsugrarpL}
\end{figure}
As for the first peak, we have $l_2^\Lambda > l_2^\SUGRA >
l_2^\RP$. Roughly speaking, for $h \simeq 0.62$, $\Omega_\MAT \simeq
0.3$, we have $l_2^\Lambda \simeq 550$, $l_2^\SUGRA \simeq 525$ and
$l_2^\RP \simeq 500$. Interestingly enough, the SUGRA QCDM model seems
to predict the correct location of the ``suggested second
peak''~\cite{MAX1a}, just in between the location predicted by the
$\Lambda$CDM model and the Ratra-Peebles QCDM model. Of course, it is
premature to conclude and only more data could allow to know whether
this is indeed the case or whether this is just a coincidence.
\par
Finally, we display the multipole moments for the $\Lambda$CDM model,
the Ratra-Peebles QCDM model and the SUGRA QCDM model in
Figs.~\ref{clL},~\ref{clrp} and~\ref{clsugra}, respectively, for the
following cosmological parameters (deduced from the previous
considerations): $h = 0.62$, $\Omega_\Lambda = 0.7$, $\Omega_\BAR =
0.0595$ and $n_\SCAL = 0.99$. The data points of COBE, BOOMERanG,
MAXIMA-1 and Saskatoon have been added to the plots for
comparison. These curves represent the predictions of each model and
special attention must be paid to third peak which is certainly one of
the next important experimental challenge. In Fig.~\ref{clsugrarpL},
we present the three curves together in order to make the comparison
easier. It should be emphasized again
that the multipole moments predicted by the SUGRA QCDM model are
unique in the sense that they do not depend on the free parameter in
the potential. From these plots, we see that the SUGRA QCDM model is,
among the three models studied here, the best fit of the MAXIMA-1
data. It is the only model for which the theoretical curve $\delta T$
versus $l$ goes through all the $1 \sigma$ error bars of this
experiment. However, we should be careful not to overestimate the
relevance of this result since uncertainties are still large, for
instance because the comparison of the calibrations of different
experiments is always a difficult task. We should also keep in mind
that $2 \sigma$ deviations are always possible. Thus, we are waiting
eagerly for the new data to see whether quintessence, and especially
SUGRA quintessence, can confirm the hints of this article and fits the
data better than the other QCDM models.

\section{conclusion}
\label{sec_conc}

The quintessence scenario provides a general framework within which
the issue of the energy density of the Universe can be tackled.  In
particular long-standing issues such as coincidence problem (and maybe
the fine-tuning problem) receive reasonnable answers for a class of
models possessing the property of tracking fields, \IE{} the evolution
of the quintessence field is driven at small redshift towards an
attractor independently of the initial conditions. In the same spirit
it seems very enticing to draw consequences of the quintessence
hypothesis on other cosmological observables, the most prominent ones
being the cosmological anisotropies. Recent measurement of the CMB
anisotropies by the BOOMERanG and MAXIMA-1 experiments give a first
indication on the location of the peaks in the CMB multipoles.  It
seems therefore topical to understand the consequence of the
quintessence hypothesis on the CMB anisotropies.
\par
In this paper we have confronted analytical methods with numerical
results.  Using the former we establish that the quintessence
perturbation are independent of the initial conditions. This is
confirmed by a full numerical computation.  This allows us to study
the CMB anisotropies. In particular we have paid particular attention
to the comparison between three possible models: the cosmological
constant model, the Ratra-Peebles and SUGRA quintessence models. We
have also compared these three models with the existing data from the
BOOMERanG and MAXIMA-1 experiments. As a rule the location of the
first peak is shifted to the right for models having an equation of
state $\omega$ closer to $-1$. This entails that the location of the
first peak for the first peak of the MAXIMA-1 data is fitted by the
SUGRA model. Similarly the location of the second peak around $l_2
\simeq 525$ as suggested by MAXIMA-1 seem to indicate that the SUGRA
model comes closer to be the best of these three models.  One of the
foreseeable challenges will be to carry out a thorough analysis of the
forthcoming data in order to distinguish these three models even more
clearly.
\par
{}From the particle physics point of view most of the quintessence
models discussed so far have neglected the crucial effects of SUSY
breaking. It may well be that the effects of SUSY breaking, on top of
necessitating a severe fine-tuning of the cosmological constant, will
induce drastic modifications in the functional form of the
quintessence potential.  It is certainly a tantalizing challenge to
include the effects of SUSY breaking within the supergravity models of
quintessence~\cite{us}.  On the other hand there exists the
possibility that the cosmological constant problem will be resolved
using ideas stemming from extra-dimension scenarios involving an
effective supersymmetry in four dimensions~\cite{Witten}. The
investigation of such models might well shed new light on the origin
of the quintessence field.

As must be clear by now the issues raised by the cosmological constant
problem, the quintessence scenario and its proper understanding within
particle physics are many-fold. The experimental results which will be
available in the near future might help in disentangling some of these
very conspicuous matters.

\end{document}